\documentclass[trackchanges]{aastex701}
\usepackage{amsmath}

\usepackage{booktabs}
\usepackage{booktabs}
\usepackage{multirow}
\usepackage{caption}
\newcommand{\kms}{km s$^{-1}$}

\begin{document}

\title{Sloshing Oscillations in coronal loops excited by successive M- and C-Class flares}

\author[orcid=0009-0002-4824-4234,sname='Paliwal']{Hitesh Paliwal}
\affiliation{Aryabhatta Research Institute of Observational Sciences (ARIES), Manora Peak, Nainital-263001, Uttarakhand, India}
\affiliation{Department of Physics, Indian Institute of Technology Roorkee, Roorkee – 247667, Uttarakhand, India}

\email[show]{hitesh.paliwal@aries.res.in}  

\author[orcid=0000-0002-0735-4501, sname='Krishna Prasad'] {S. Krishna Prasad}
\affiliation{Aryabhatta Research Institute of Observational Sciences (ARIES), Manora Peak, Nainital-263001, Uttarakhand, India}
\email{krishna.prasad@aries.res.in}
\author[orcid=0000-0002-5359-5592, sname='Sunil Krishna'] {M. V. Sunil Krishna}
\affiliation{Department of Physics, Indian Institute of Technology Roorkee, Roorkee – 247667, Uttarakhand, India} 
\affiliation{Centre for Space Science and Technology, Indian Institute of Technology Roorkee, Roorkee –247667, Uttarakhand,
India}
\email{mv.sunilkrishna@ph.iitr.ac.in}


\begin{abstract}
Slow magnetoacoustic waves in hot coronal loops have remained a topic of considerable interest and debate over the past two decades. The periodic back-and-forth motion of plasma within a coronal loop, often initiated by a flare, is commonly referred to as sloshing oscillation. In the present study, we report unprecedented observations of sloshing oscillations in coronal loops excited by successive M- and C-class flares, using data from the Atmospheric Imaging Assembly (AIA) onboard the Solar Dynamics Observatory (SDO). A total of fifteen oscillation events were identified within seven distinct coronal loops, providing the rare opportunity to evaluate the influence of flare strength on the characteristics of the oscillations. Based on the appearance of the oscillations, their properties were extracted mainly from the AIA 131 and 94 \AA\ channels. Additionally, we estimate the deprojected length of each loop by assuming a semi-circular geometry. Our results indicate considerable changes in the properties of oscillations from one flare to another, suggesting the role of individual flares in shaping the local physical conditions. The plasma temperature estimated from the loop length and oscillation period ranges from 9 to 31 MK. Additionally, we find that the damping times are not always longer in the colder 94 \AA\ channel as previously observed. By combining the results obtained from all events, we study the inter-dependences between various parameters, including oscillation period, damping time, loop length, and plasma temperature, and discuss these results in the context of the theory of slow waves.
\end{abstract}

\keywords{Sun: Magnetohydrodynamics (MHD) --- Sun: corona ---  Flare --- 
 oscillations --- waves}


\section{Introduction} 
\label{intro}
The solar corona hosts different kinds of magnetohydrodynamic (MHD) waves/oscillations, which are interesting to study because they are possible candidates for heating the corona \citep{2020SSRv..216..140V} and are also very important for probing the corona using coronal seismology \citep{2000SoPh..193..139R, 2012RSPTA.370.3193D, 2014SoPh..289.3233L, 2018AdSpR..61..655A}. The slow magneto-acoustic wave is one of the MHD waves, which has been extensively studied in coronal loops previously (see \citealt{2021SSRv..217...34W} and \citealt{Banerjee2021} for latest reviews). The standing versions of these waves are often observed in hot loops associated with flares.

 The first detection of compressive waves in hot coronal loops was reported by \cite{Wang2002}, using Doppler shift in Fe XIX (6.3 MK) and Fe XX (8 MK) spectral lines observed with the Solar Ultraviolet Measurements of Emitted Radiation (SUMER) instrument aboard the Solar and Heliospheric Observatory (SoHO). A study of 54 Doppler-shift oscillations in 27 flare-like events showed that the oscillation periods range from 7 to 31 minutes, with damping times between 6 and 37 minutes \citep{Wang2003}. The maximum Doppler shift velocities were found to be in the range of 100 -- 300 \kms\ \citep{Wang2005}. The speed is found to be nearly equal to the local sound speed, and in some cases, a quarter-period phase lag is observed between the Doppler velocity and intensity fluctuations \citep{2003A&A...402L..17W}. These characteristics suggest that the oscillations can be considered as standing slow magnetoacoustic waves. In the study by \cite{Wang2003}, only a few events were found to be associated with flares; most of the oscillations were not directly related to any flare activity. Since these were all limb events, it is possible that flares occurring on the far side of the solar disk could have triggered them. Furthermore, some of these oscillations were accompanied by brightenings at one of the footpoints \citep{Wang2005}, suggesting that microflares could also be responsible for their trigger. 
 
Later, \cite{2005ApJ...620L..67M} and \cite{2006ApJ...639..484M} studied coronal loop oscillations during flares using Doppler shift measurements obtained with the Bragg Crystal Spectrometer (BCS) onboard Yohkoh. The average period of the BCS oscillations was 5.5 $\pm$ 2.7 minutes, and the average damping time was 5.0 $\pm$ 2.5 minutes \citep{2006ApJ...639..484M}. The BCS oscillations also exhibit rapid damping similar to the SUMER oscillations, although their periods are shorter. The shorter periods may plausibly be associated with the hotter loops observed by BCS as it detects emission from plasma at very high temperatures ($ \sim 12\ \text{MK} $). A similar study was later carried out using the EUV Imaging Spectrometer (EIS) onboard the Hinode satellite \citep{2008ApJ...681L..41M}. In this case, the oscillation period was found to be approximately 35 minutes, with a decay time of about 43 minutes. Later, \cite{2012ApJ...756L..36K} reported the presence of slow waves in a coronal loop during an M1.6 flare using the 17 GHz channel of the Nobeyama Radioheliograph. The oscillations exhibited rapid damping, with a period of 12.6 minutes and a decay time of 15 minutes, consistent with previous studies conducted on hot coronal loops.

Overall, the damping times of the oscillations were comparable to their oscillation periods, indicating that they exhibit rapid damping. The best-fit power-law scaling of the damping time with the oscillation period yields an exponent of 0.96 $\pm$ 0.18 \citep{Wang2005}. A 1D MHD simulation by \cite{Ofman2002} shows that thermal conduction is the primary damping mechanism for these oscillations. Recent studies also suggest that compressive viscosity \citep{Wang2015, Wang2018} and heating-cooling misbalance \citep{Nakariakov2017, 2019A&A...628A.133K} could also play a significant role in their damping under different physical conditions in coronal loops. Various numerical and theoretical studies have been conducted to understand the properties of slow magnetoacoustic waves, including their driving mechanisms \citep{2005A&A...436..701S, 2009AnGeo..27.3899S, 2012ApJ...754..111O}, damping mechanisms \citep{Ofman2002, 2008A&A...483..301B, 2008ApJ...685.1286V, 2013A&A...553A..23R, 2014ApJ...786...36A, Wang2018}, and other related characteristics \citep{2004A&A...414L..25N, 2006A&A...446.1151N, 2015ApJ...807...98Y, 2015ApJ...813...33F}. Despite these numerous studies, the cause of the rapid excitation of standing slow magnetoacoustic waves in hot loops and their damping are still not well understood.

Another type of longitudinal oscillation in hot coronal loops was first reported by \cite{Berghmans2001} using observations from the Soft X-ray Telescope (SXT) aboard Yohkoh. In this event, which occurred in a loop with a temperature of about 4$\--$5 MK, a plasma perturbation was seen to bounce back and forth between the loop footpoints with a gradually decreasing amplitude. The oscillation was triggered by a microflare near one of the loop footpoints. The measured phase speed was approximately 310 \kms, comparable to the local sound speed of 300--350 \kms, leading the authors to interpret these disturbances as propagating slow magnetoacoustic waves. Similar, but much clearer,  oscillatory behavior was later observed by \cite{Kumar2013} in a hotter coronal loop with a temperature of about 8$\--$10 MK using high resolution data from the Atmospheric Imaging Assembly (AIA) onboard the Solar Dynamics Observatory (SDO). In this case, the oscillations were triggered by a C-class flare near one of the loop footpoints and showed rapid damping, with a period of about 10 minutes and a decay time of nearly 7 minutes. The estimated phase speed of 460$\--$510 \kms\ was again comparable to the local sound speed of 430$\--$480 \kms. These properties were similar to those of oscillations observed with SUMER, except for the fact that the oscillations in this case were found to be propagating between the foot points of the loop. Indeed, \citet{Kumar2013} identified them as reflected, propagating slow magnetoacoustic waves. Subsequently, \cite{Reale2016} referred to these oscillations as sloshing oscillations.

During the onset of sloshing oscillations, in some events, brightenings at both footpoints have been reported \citep{Wang2015, Kumar2015}. In fact, the study by \cite{Wang2015} reported a distinct case of a hot coronal loop that provided clear evidence of standing waves. However, both studies suggested that a fan–spine magnetic topology with reconnection at the fan could account for the simultaneous brightenings at both footpoints.  The general properties of the oscillations reported by \cite{Wang2015} were similar to those of sloshing oscillations, except that they did not exhibit a back-and-forth plasma motion. Instead, they displayed an antiphase variation of plasma between the footpoints, which is characteristic of standing slow magnetoacoustic waves. The study further demonstrated that thermal conduction was suppressed, while compressive viscosity was enhanced by more than an order of magnitude in that loop.  Using one-dimensional nonlinear MHD simulations, \cite{Wang2018} further demonstrated that sloshing oscillations can transform into standing oscillations after a few cycles, depending on the physical conditions of the loop. Later, \cite{KrishnaPrasad2021} observationally demonstrated that sloshing oscillations indeed evolve into standing waves after a few cycles. These authors utilized the same AIA/SDO dataset analyzed by \cite{Kumar2013} but observed in the AIA 94 \AA\ channel. The oscillation period and damping time during the sloshing phase were consistent with those reported by \cite{Kumar2013}, whereas both parameters were higher during the standing phase, likely due to a decrease in temperature during the transition from sloshing to standing oscillations. The authors also found that the damping time in the 94 \AA\ channel was approximately twice that in the 131 \AA\ channel during the sloshing phase, indicating a multi-thermal nature of the plasma. More recently, \cite{2024SoPh..299..163K} investigated the properties of sloshing oscillations triggered by M-class flares using the 131 \AA\ channel of AIA/SDO. They identified a total of 16 events, although six of them do not display a clear oscillatory pattern. The authors found that the overall oscillation periods range from 150 s to 1325 s, with an average propagation speed of about 500 \kms. Importantly, some of the periods detected in this study are much shorter (3-5 min range) than those reported in previous studies \citep{Kumar2013, Kumar2015, Wang2015, Mandal2016} where C-class flare is the primary trigger. 

In the present study, we report unique observations of coronal loops in which sloshing oscillations were triggered by successive M- and C-class flares. We analyze their properties and investigate the scaling relationships between the oscillation period, damping time, and loop length. Section~\ref{observation} provides an overview of the selected events, followed by a description of the data analysis in Section~\ref{analysis}. The final results and discussion are presented in Section~\ref{results}, and our main conclusions are summarized in Section~\ref{conclusion}.


\section{Observations}
\label{observation}
To study sloshing oscillations in flare-associated coronal loops, we utilised the data from AIA \citep{AIA} onboard SDO \citep{SDO}. The multi-wavelength coverage of AIA, with various filters, makes it sensitive to plasma across a wide range of temperatures, from the chromosphere to the flaring corona. Therefore, AIA is particularly well-suited for studying dynamic phenomena in the solar atmosphere. The initial search for sloshing oscillations was conducted using JHelioviewer\footnote{\url{https://www.jhelioviewer.org/}}. Since these oscillations are most commonly observed in the AIA 131 \AA\ channel \citep{Kumar2013, Kumar2015, Wang2015, Mandal2016, KrishnaPrasad2021, 2024SoPh..299..163K}, we examined low-resolution animations in this channel and selected events where the plasma exhibited a clear back-and-forth (sloshing) motion within the same coronal loop during M- and C-class flares. A total of 15 events were identified within 7 distinct coronal loops as listed in Table~\ref{table}, where the start time of each of these events is provided. The cutout data corresponding to all events were then downloaded and processed using a suite of IDL routines developed by Rob Rutten\footnote{\url{https://robrutten.nl/rridl/sdolib/dircontent.html}}. This pipeline co-aligns and de-rotates the images, producing level~1.5 data, which were used for further analysis. The plate scale and temporal cadence of the level~1.5 AIA data are 0.6~arcsec and 12~s, respectively. Variations in exposure time during flares were corrected by normalizing each image by its exposure duration.  

The class and timing of each flare in our event list were identified using the soft X-ray light curves obtained from the X-ray Sensor (XRS) onboard the Geostationary Operational Environmental Satellites (GOES) \citep{ GOES, XRS/GOES-R}. The GOES XRS measures the integrated solar soft X-ray flux in two wavelength bands (0.5$\--$4 \AA\ and 1$\--$8 \AA) with a cadence of 2 s, providing a reliable classification of flare strength according to the peak flux in the 1$\--$8 \AA\ channel. The flare catalog used for this study is accessible via the Space Weather Live database\footnote{\url{https://www.spaceweatherlive.com/en/solar-activity/solar-flares.html}}, which compiles GOES event data and metadata for all recorded solar flares.

\section{Analysis}
\label{analysis}
As the loops are clearly identifiable only in the AIA 131~\AA\ and 94~\AA\ channels, we restrict our analysis to these two passbands. Here, we present the detailed analysis procedures for a sample case corresponding to loop 1 (see Table \ref{table}), but all the remaining loops are analyzed following the same procedure, and corresponding results are presented in Appendix~\ref{events}.

\begin{figure}
    \centering
    \includegraphics[width=1\textwidth]{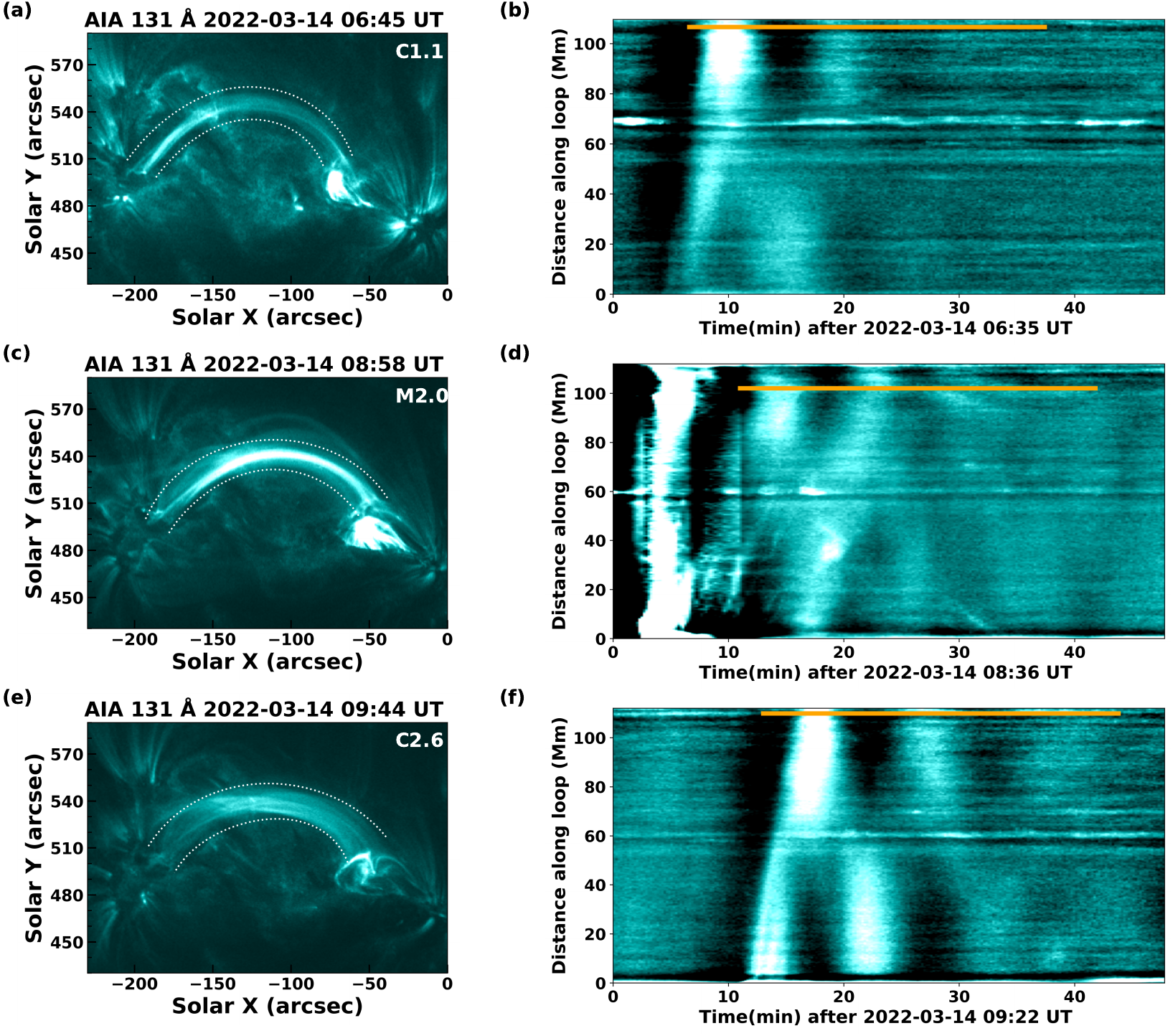}
   \caption{(a) Map of the loop structure corresponding to Event a in loop 1 listed in Table~\ref{table}. The chosen loop boundaries are marked by white dotted lines, and the corresponding flare class is indicated in the top-right corner. (b) Time-distance plot for the loop selected in panel~(a), where the solid orange line marks the position from which the time series was extracted for further analysis. Panels~(c) and~(e) are similar to panel~(a) but correspond to events b and c in loop 1 in Table~\ref{table}, respectively. Panels~(d) and~(f) show the corresponding time-distance plots for the loop in panels~(c) and~(e), respectively. An animation depicting the evolution of all three events is available online for both 131 and 94 \AA\ channels.}

    \label{xt}
\end{figure}

In order to visualize the sloshing oscillations in a loop, we first created time-distance maps by manually selecting the loop boundaries, integrating the intensities across the loop width from each frame, and stacking them together. The traced loop and the constructed time-distance map for all three events corresponding to loop 1 are presented in Figure~\ref{xt}. The associated animation displays the time evolution of the loop. The bright, slanted ridges alternating forward and backward in the time-distance plots reveal a plasma perturbation moving back and forth, indicative of sloshing oscillations within the loop. To enhance the visibility of the oscillations, we detrended and normalized the intensities using a running average with a window size longer than the oscillation period. Detrending was performed solely to improve the visualization of the oscillations, whereas the analysis to extract the wave properties was carried out using the original timeseries. 

Subsequently, to determine the oscillation properties, we extracted the time series from a fixed spatial location near one of the loop footpoints (indicated by the orange line in Figure~\ref{xt}). This choice was made deliberately because locations away from the foot point along the loop are projected onto the solar disk. As a consequence, selecting an arbitrary position along the loop may lead to differences in the apparent oscillation period \citep{KrishnaPrasad2021}. The resulting light curve was modeled following the approach of \cite{KrishnaPrasad2021}, with certain modifications to account for background variations. More specifically, instead of applying the model to a background-subtracted light curve, we included a second-order polynomial term that represents a slowly varying background trend in the model itself. This way, we can avoid any bias that may be introduced by manually selecting minima for parabolic background removal (as was done in earlier works) and thus improve the robustness of the fit. The final fitted function can therefore be expressed as

\begin{figure}
    \centering
    \includegraphics[width=1\textwidth]{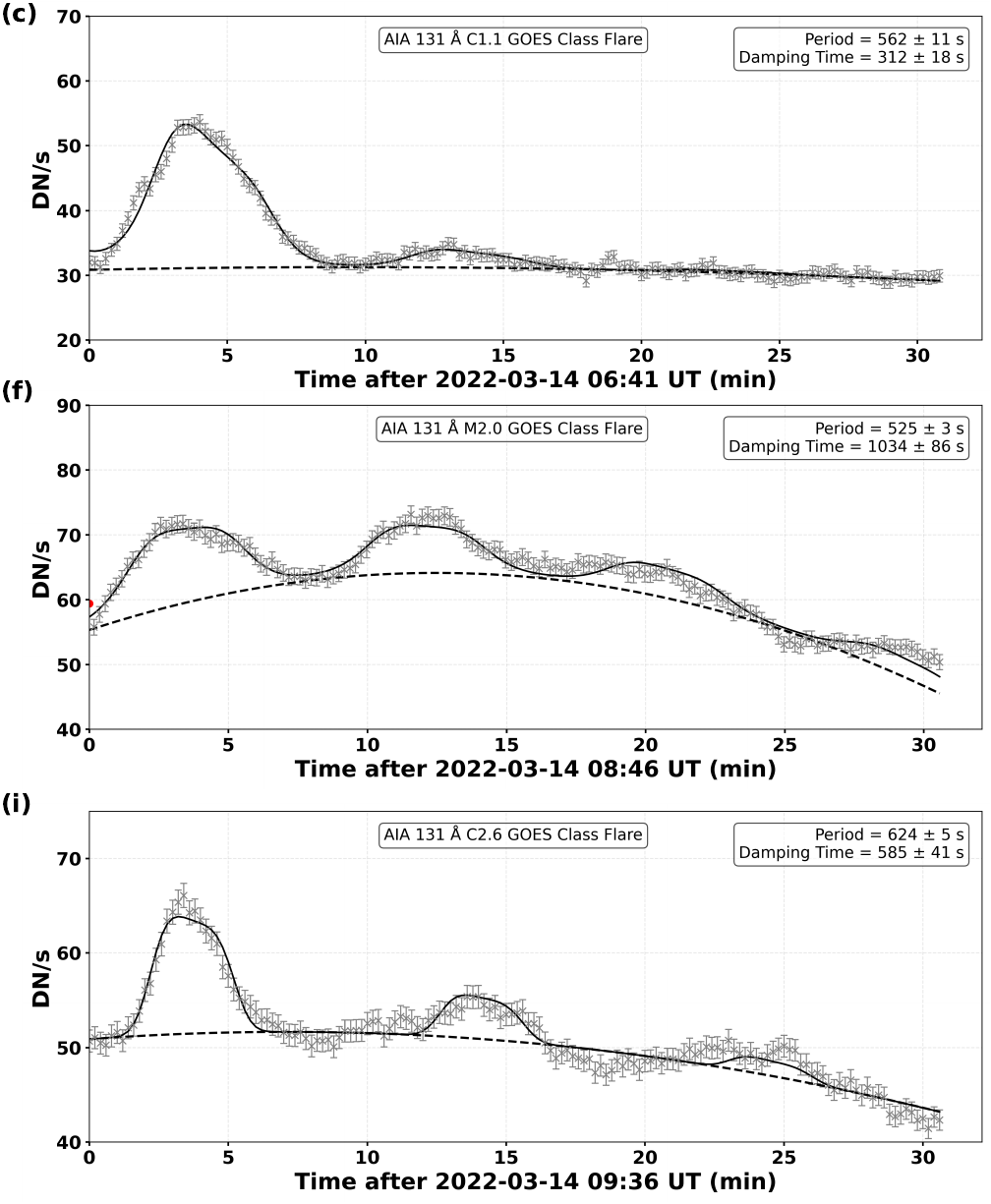}
  \caption{ Extraction of oscillation parameters. The symbols with error bars represent data points with the nominal exposure time extracted from the original time--distance maps at the locations marked by the orange solid lines in Figure~\ref{xt}. Data points shown by red markers correspond to frames with reduced exposure times and were excluded from the fitting procedure. The solid black line represents the best fit to the data using the model described by Equation~\ref{eq1}, while the black dashed line denotes a second-order polynomial background included in the fit. The final best-fit parameters obtained are listed in the plot.}

    \label{fitting}
\end{figure}

\begin{equation}
I(t) = A_0 \exp\left(-\frac{t}{\tau}\right) \exp\left(-\frac{ x_0^2}{\sigma^2}\right) + B_0 \, t^2 + C_0 \, t + D
\label{eq1}
\end{equation}

where,
\begin{equation*}
x_0 = \frac{L}{2} \left[1 - \cos\left(\frac{2\pi t}{P} + \phi\right)\right]
\end{equation*}

Here, $t$ is time, $L$ is the length of the structure, $P$ is the oscillation period, $\tau$ is the damping time, and $\sigma$ is the spatial width of the perturbation. $A_{0}$, $B_{0}$, $C_{0}$, $D$, and $\phi$ are appropriate constants. The light curves extracted along the orange solid lines in the time-distance plots shown in Figure~\ref{xt}, together with the corresponding best-fit curves obtained by applying the model described in Equation~\ref{eq1} using chi-square minimization, are presented in Figure~\ref{fitting}. The fitting was performed using a non-linear least-squares minimization algorithm based on the Levenberg-Marquardt method, implemented via the LMFIT library in Python. This approach yielded best-fit estimates of the oscillation parameters along with their associated uncertainties. The uncertainties in our observed data are calculated following \cite{Yuan2012}, using

{\small
\begin{equation}
\sigma_{\text{131}}(F) \approx \sqrt{2.392 + 0.06 F + 8.1 \times 10^{-7} F^2}\; (\text{DN})
\label{error1}
\end{equation}
}

{\small
\begin{equation}
\sigma_{\text{94}}(F) \approx \sqrt{2.299 + 0.058 F + 8.1 \times 10^{-7} F^2}\; (\text{DN})
\label{error2}
\end{equation}
} 

where F is the observed counts (in DN) per pixel in level 1 data. Equation~\ref{error1} represents the uncertainty in AIA 131 \AA\ whereas equation~\ref{error2} gives that in 94 \AA\ in units of data numbers (DN). These expressions include photon Poisson noise, dark-current noise, electronic readout noise, digitization noise, compression noise, subtraction noise, and noise introduced during the removal of spikes from the images. Uncertainties introduced during level 1 to level 1.5 data processing are not considered, as the irreversible nature of these steps prevents a reliable evaluation of noise propagation. It should be noted that these uncertainties are valid only when estimated at the original observed counts (not the normalised values) and for proper comparison of intensities along with their errors across time, there should identical exposures. However, during the peak phase of flares, particularly for energetic M- and X-class events, the AIA automatic exposure control often reduces the exposure time of alternate frames to prevent detector saturation in specific channels. To account for this, we normalise the intensities by corresponding exposure times and propagate the computed uncertainties. Additionally, to avoid the influence of nonlinear dependence in the response of the detector, the frames with reduced exposure times were excluded from the final fitting and highlighted with red markers in Figure~\ref{fitting} and all the corresponding figures presented in Appendix~\ref{events}. The oscillation period and damping time thus obtained for all the events are listed in Table \ref{table}. In a few cases, the oscillations are not clearly visible in either the 131~\AA\ or 94~\AA\ channels, and therefore, the corresponding parameters are left blank. 

In addition to the oscillation parameters, the lengths of the respective coronal loops were estimated by assuming a semi-circular geometry that projects as an ellipse on the solar surface. To achieve this, for each flare event, we selected a reference frame in which the target loop was clearly visible. In this frame, the loop footpoints were manually identified, and multiple points along the loop were selected to trace its path. These points were then fitted using the model of \citet{2002SoPh..206...99A}, which deprojects the coronal loop using two free parameters: the height of the loop center ($h_0$) above the solar surface and the inclination angle ($\theta$) between the loop plane and the local vertical. The loop lengths obtained from the best-fit model are listed in Table \ref{table} for all 7 loops. Note that the loop length did not change between the events, so only a single value is provided for each distinct loop analysed. The traced locations and the final fitted curve corresponding to loop 1 are shown as a dark curve in Figure~\ref{loop}b in appendix~\ref{app}. Note that the uncertainties quoted for the loop lengths reflect only the statistical goodness of the model fit and do not account for systematic uncertainties associated with the assumption of a semi-circular geometry. Additionally, since the reconstruction is based on a projected two-dimensional image and no direct three-dimensional information is available, the true uncertainty in the loop length may be substantially larger.

It may be seen that the oscillation periods are generally shorter during M-class flares in comparison with those during C-class flares, although the difference is marginal in most cases. Since a hotter plasma would naturally support faster propagation speeds, resulting in shorter periods, it is possible that the loop plasma is hotter during M-class flares. To investigate this further, we attempted to retrieve temperature information using the Differential Emission Measure (DEM) technique, following \citet{2012A&A...539A.146H}. However, we find that the peak temperature values are clustered around 10-12 MK, and their time dependence often displays a saturation near the peak value, especially during M-class flares. This indicates a limitation of our DEM analysis solely based on AIA EUV channel data, which does not include very high temperature emission. Therefore, we proceed to estimate plasma temperature using an alternative approach. First, we calculate the phase speed $v$ from the obtained loop length $L$ and oscillation period $P$ using $v$ = $2L$/$P$. One may note that a similar expression is generally used for fundamental standing mode. Although, the sloshing oscillations do not behave like standing waves (in contrast, they display propagation from one foot point to another), here period (P) is the time required for the perturbation to travel from one footpoint to the other and return to the original footpoint after reflection. Thus, during one period, the perturbation propagates a total distance of approximately (2L), yielding the same dependence for the phase speed. Next, by assuming the phase speed is equivalent to the local sound speed, one could write its relation with plasma temperature $T$ as 
\begin{equation}
v = \sqrt\frac{\gamma k_B T}{\mu m_p}
\label{eq:speed}
\end{equation}
where \( \gamma = 1.67 \) is the adiabatic index, \( k_B \) is the Boltzmann constant, \( \mu = 0.6 \) is the mean molecular weight of particles in the corona, and \( m_p \) is the mass of a proton. By substituting standard values for all the constants, the value of $T$ can be calculated following $T=(v/152)^2$, with the value of $v$ expressed in \kms and $T$ in MK. The obtained phase speed and plasma temperature values for all loops are listed in Table \ref{table}. Note that an average value of oscillation period found in 131~\AA\ or 94~\AA\ channels is used in these calculations. Although the plasma temperatures estimated here cannot be used to comprehend the shorter periods during M-class flares (since the period value is used in their calculation), the substantially larger values obtained highlight the need for inclusion of hotter plasma emission in DEM calculations.

\section{Results and Discussion}
\label{results}

\renewcommand{\arraystretch}{1.4}
\setlength{\tabcolsep}{2.9pt}
\begin{table*}
\caption{The parameters extracted for all 15 sloshing oscillation events analyzed. These include the start time,  GOES class of the flare, loop length $L$, damping time $\tau$, oscillation period $P$, phase speed $v$, and plasma temperature $T$ (obtained from phase speed), analyzed in AIA 131~\AA\ and 94~\AA\ channels.}
\centering

\begin{tabular}{ccccccccccc}
\toprule
\textbf{Loop} & \textbf{Event} & \textbf{GOES} & \textbf{Coordinate} &
\textbf{Start time} & \textbf{$L$} &
\textbf{AIA} &
\textbf{$\tau$} &
\textbf{$P$} &
\textbf{$v$} &
\textbf{$T$} \\
\textbf{No.} & \textbf{No.} & \textbf{Class} & & \textbf{} &
\textbf{(Mm)} & \textbf{Channel} &
\bf{(s)} & \bf{(s)} & \bf{(\kms)} & \bf{(MK)} \\
\midrule
\label{table}

\multirow{6}{*}{1} &
\multirow{2}{*}{a} &
\multirow{2}{*}{C1.1} &
\multirow{2}{*}{N23E04} &
\multirow{2}{*}{2022-03-14T06:38} &
\multirow{6}{*}{179.5 $\pm$ 2.7} &
131 \AA & 312 $\pm$ 18 & 562 $\pm$ 11 &\multirow{2}{*}{628 $\pm$ 10} &\multirow{2}{*}{17.1 $\pm$ 0.5} \\
& & & & & & 94 \AA & 703 $\pm$ 44 & 581 $\pm$ 5  && \\

& \multirow{2}{*}{b} & \multirow{2}{*}{M2.0} &
\multirow{2}{*}{N23E03} &
\multirow{2}{*}{2022-03-14T08:29} &
& 131 \AA & 1034 $\pm$ 86 & 525 $\pm$ 3 & \multirow{2}{*}{650 $\pm$ 12} & \multirow{2}{*}{18.3 $\pm$ 0.6}\\
& & & & & & 94 \AA & 387 $\pm$ 52 & 582 $\pm$ 17  & & \\

& \multirow{2}{*}{c} & \multirow{2}{*}{C2.6} &
\multirow{2}{*}{N23E03} &
\multirow{2}{*}{2022-03-14T09:45} &
& 131 \AA & 585 $\pm$ 41 & 624 $\pm$ 5 &\multirow{2}{*}{569 $\pm$ 7} & \multirow{2}{*}{14.0 $\pm$ 0.3}\\
& & & & & & 94 \AA & 912 $\pm$ 74 & 637 $\pm$ 4  & & \\
\midrule

\multirow{4}{*}{2} &
\multirow{2}{*}{a} &
\multirow{2}{*}{C6.5} &
\multirow{2}{*}{N28W15} &
\multirow{2}{*}{2025-09-06T18:49} &
\multirow{4}{*}{210.0 $\pm$ 7.2} &
131 \AA & 270 $\pm$ 13 & 492 $\pm$ 7 & \multirow{2}{*}{822 $\pm$ 21} & \multirow{2}{*}{29.2 $\pm$ 1.5} \\
& & & & & & 94 \AA & 636 $\pm$ 36 & 532 $\pm$ 4  & & \\

& \multirow{2}{*}{b} &
\multirow{2}{*}{M1.2} &
\multirow{2}{*}{N28W16} &
\multirow{2}{*}{2025-09-06T22:07} &
& 131 \AA & 215 $\pm$ 12 & 465 $\pm$ 9 &\multirow{2}{*}{853 $\pm$ 23}& \multirow{2}{*}{31.5 $\pm$ 1.7} \\
& & & & & & 94 \AA & 341 $\pm$ 30 & 523 $\pm$ 8  & & \\
\midrule

\multirow{4}{*}{3} &
\multirow{2}{*}{a} &
\multirow{2}{*}{C1.4} &
\multirow{2}{*}{S28W13} &
\multirow{2}{*}{2022-06-16T02:03} &
\multirow{4}{*}{76.9 $\pm$ 3.2} &
131 \AA & 129 $\pm$ 13 & 320 $\pm$ 12 & \multirow{2}{*}{482 $\pm$ 17} & \multirow{2}{*}{10.1 $\pm$ 0.7} \\
& & & & & & 94 \AA & 182 $\pm$ 17 & 318 $\pm$ 5  & & \\

& \multirow{2}{*}{b} &
\multirow{2}{*}{M1.6} &
\multirow{2}{*}{S28W14} &
\multirow{2}{*}{2022-06-16T03:52} &
& 131 \AA & 191 $\pm$ 25 & 306 $\pm$ 10 & \multirow{2}{*}{504 $\pm$ 17}& \multirow{2}{*}{11.0 $\pm$ 0.8} \\
& & & & & & 94 \AA & 350 $\pm$ 37 & 304 $\pm$ 4  &  &\\
\midrule

\multirow{4}{*}{4} &
\multirow{2}{*}{a} &
\multirow{2}{*}{M2.6} &
\multirow{2}{*}{N12E18} &
\multirow{2}{*}{2014-02-02T06:25} &
\multirow{4}{*}{104.7 $\pm$ 6.5} &
131 \AA & 415 $\pm$ 63 & 331 $\pm$ 6 &  \multirow{2}{*}{639 $\pm$ 29} & \multirow{2}{*}{17.7 $\pm$ 1.6} \\
& & & & & & 94 \AA & 413 $\pm$ 46 & 324 $\pm$ 4  &  &  \\

& \multirow{2}{*}{b} &
\multirow{2}{*}{C5.2} &
\multirow{2}{*}{N12E16} &
\multirow{2}{*}{2014-02-02T11:20} &
& 131 \AA & 333 $\pm$ 24 & 454 $\pm$ 6 &  \multirow{2}{*}{446 $\pm$ 20} & \multirow{2}{*}{8.6 $\pm$ 0.8} \\
& & & & & & 94 \AA & 247 $\pm$ 30 & 485 $\pm$ 13  & &  \\
\midrule

\multirow{4}{*}{5} &
\multirow{2}{*}{a} &
\multirow{2}{*}{C1.4} &
\multirow{2}{*}{S28W13} &
\multirow{2}{*}{2022-06-16T02:03} &
\multirow{4}{*}{160.9 $\pm$ 8.3} &
131 \AA & --- & --- &  \multirow{2}{*}{522 $\pm$ 29} & \multirow{2}{*}{11.8 $\pm$ 1.3}\\
& & & & & & 94 \AA & 1261 $\pm$ 381 & 616 $\pm$ 14  & &  \\

& \multirow{2}{*}{b} &
\multirow{2}{*}{M1.6} &
\multirow{2}{*}{S28W14} &
\multirow{2}{*}{2022-06-16T03:52} &
& 131 \AA & 313 $\pm$ 42 & 460 $\pm$ 15 &  \multirow{2}{*}{675 $\pm$ 29} & \multirow{2}{*}{19.7 $\pm$ 1.7} \\
& & & & & & 94 \AA & 1660 $\pm$ 699 & 495 $\pm$ 14  &  &  \\
\midrule

\multirow{4}{*}{6} &
\multirow{2}{*}{a} &
\multirow{2}{*}{C1.4} &
\multirow{2}{*}{S28W13} &
\multirow{2}{*}{2022-06-16T02:03} &
\multirow{4}{*}{105.5 $\pm$ 3.7} &
131 \AA & 304 $\pm$ 27 & 372 $\pm$ 6 &\multirow{2}{*}{548 $\pm$ 16} & \multirow{2}{*}{13.0 $\pm$ 0.8} \\
& & & & & & 94 \AA & 655 $\pm$ 239  & 399 $\pm$ 13 & & \\

& \multirow{2}{*}{b} &
\multirow{2}{*}{M1.6} &
\multirow{2}{*}{S28W14} &
\multirow{2}{*}{2022-06-16T03:52} &
& 131 \AA & 624 $\pm$ 59 & 459 $\pm$ 5 & \multirow{2}{*}{460 $\pm$ 17} & \multirow{2}{*}{9.1 $\pm$ 0.7} \\
& & & & & & 94 \AA & --- & ---  &  &  \\
\midrule

\multirow{4}{*}{7} &
\multirow{2}{*}{a} &
\multirow{2}{*}{M2.0} &
\multirow{2}{*}{N22W04} &
\multirow{2}{*}{2022-03-15T22:40} &
\multirow{4}{*}{180.7 $\pm$ 7.0} &
131 \AA & 450 $\pm$ 59 & 536 $\pm$ 16 & \multirow{2}{*}{674 $\pm$ 33} &\multirow{2}{*}{19.7 $\pm$ 1.9} \\
& & & & & & 94 \AA & --- & ---  & & \\

& \multirow{2}{*}{b} &
\multirow{2}{*}{C6.7} &
\multirow{2}{*}{N28W05} &
\multirow{2}{*}{2022-03-15T23:20} &
& 131 \AA & 370 $\pm$ 30 & 558 $\pm$ 12 & \multirow{2}{*}{648 $\pm$ 29} & \multirow{2}{*}{18.2 $\pm$ 1.6} \\
& & & & & & 94 \AA & --- & ---  & & \\
\bottomrule

\end{tabular}
\end{table*}

\begin{figure}
    \centering
    \includegraphics[width=1\textwidth]{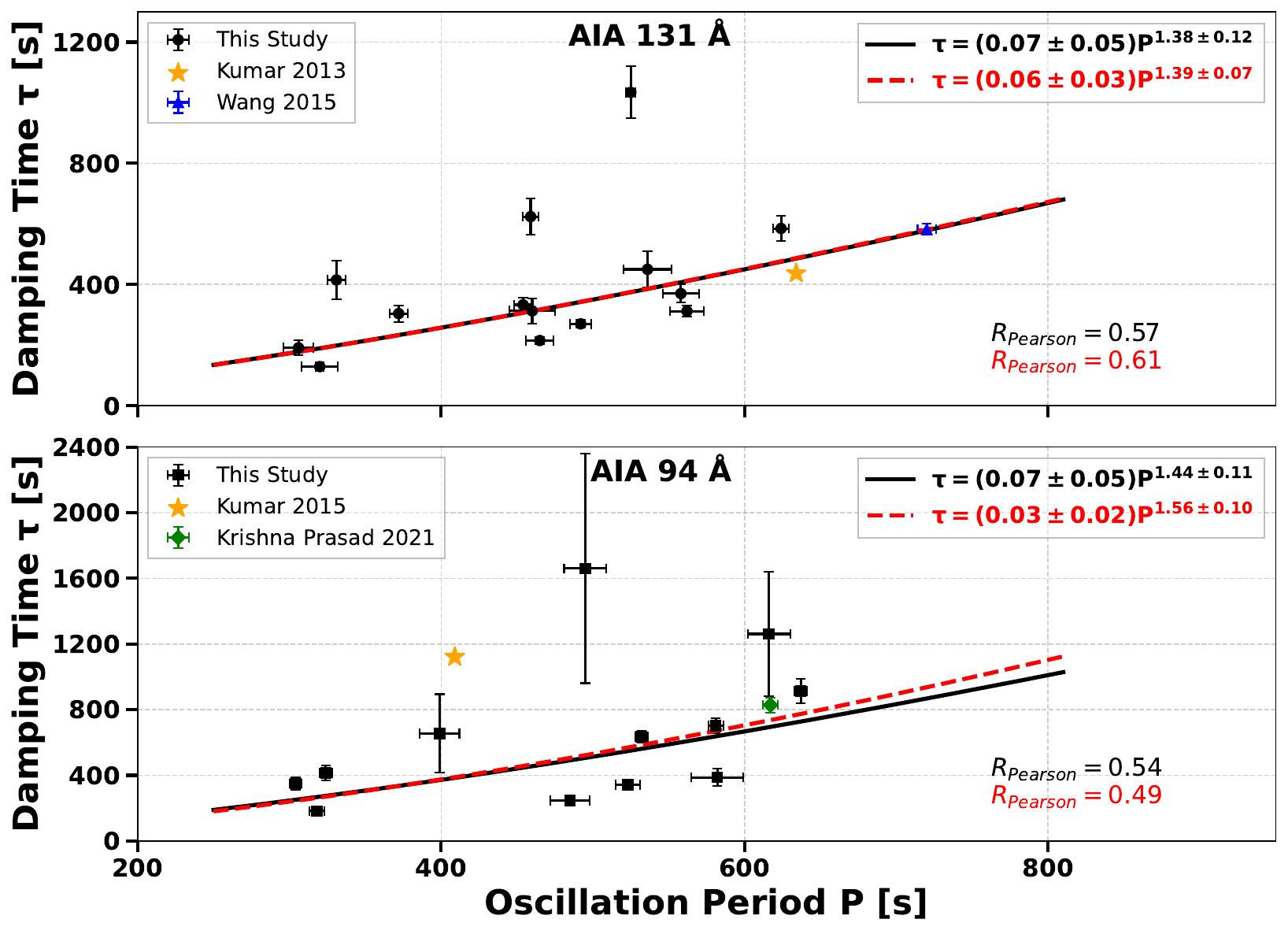}
\caption{The dependence of damping time on oscillation period in the AIA 131~\AA\ and 94 \AA\ channels. The filled black circles/squares represent the data from this work, and the color symbols represent the data from previous studies \citep{Kumar2013, Wang2015, Kumar2015, KrishnaPrasad2021} as listed in the legend. The black solid line represents the best fit to our data, while the red dashed line shows the best fit obtained when including results from previous studies. The final fitted relation and the Pearson correlation coefficient, for each case, are listed in the plot with respective colors.}
    \label{tau_vs_P}
\end{figure}

A total of 15 oscillation events observed in 7 distinct loops are analysed in this study. The extracted parameters, the loop length, oscillation period, damping time, phase speed, and plasma temperature for all 15 cases are provided in Table~\ref{table}. As noted earlier, sloshing oscillations were not detected in the 94 \AA\ channel in three cases (event 6b, 7a, and 7b) and in the 131 \AA\ channel in one case (event 5a). Specifically in the case of 5a, only a single bright ridge was observed in the time-distance map of the 131 \AA\ channel, prohibiting us from extracting the oscillation parameters. All the other cases display a clear oscillation pattern. 

Overall, the lengths of the seven distinct loops analysed lie within the range of 77-210 Mm. The obtained oscillation periods span from 304 to 637 s, while the damping times vary from 129 to 1660 s. The oscillation phase speeds range from 446 to 842 \kms, implying a corresponding change in plasma temperature from 9 to 31 MK. Prior to \citet{2024SoPh..299..163K}, who studied sloshing oscillations exclusively due to M-class flares, the oscillation periods found were $\approx$600 s or longer. Here, we observe shorter periods in most cases, even in the oscillations triggered by C-class flares. The damping times are not necessarily comparable to oscillation periods as reported earlier, but rather display much longer (e.g., events 1b, 5a, 5b, 6a) or shorter values (e.g., events 1a, 2a, 2b, 3a, 3b) in several cases. Both the period and damping time vary substantially from flare to flare, even within the same loop. Since the oscillation properties are expected to be dependent on local physical conditions, this behavior suggests the role of each flare in shaping the local conditions. For instance, the plasma temperature is, in general, higher during M-class flares, except in the case of loop 6. However, when all the events are combined, the strongest flare (event 4a) and the weakest (event 1a) yield similar plasma temperatures. This may suggest the role of additional parameters (e.g., plasma density, loop length) in determining the resultant temperature.

Regarding the physical mechanism behind the damping of sloshing oscillations, thermal conduction has been proposed as a primary mechanism in several previous studies \citep{Ofman2002, KrishnaPrasad2021}, although compressive viscosity has been shown to dominate under certain conditions \citep{Wang2015, Wang2018}. Thus, the precise mechanism responsible for the damping of these waves remains an active area of research. By comparing the damping times obtained in the 131 \AA\ and 94 \AA\ channels, it may be seen that in some events (1b, 4a, 4b), the damping time in 94~\AA\ is shorter than that in 131~\AA. This behavior is inconsistent with the results of \citet{KrishnaPrasad2021}, who reported a larger damping time in the 94~\AA\ channel relative to the 131~\AA\ channel owing to the lower plasma temperature. This implies that the dominant damping mechanism of sloshing oscillations may vary from loop to loop and is not governed solely by thermal conduction.

Furthermore, a visual comparison of time-distance maps across different events reveals that in about 6 cases (events 1c, 2a, 2b, 3a, 3b, and 6a), the sloshing oscillations exhibit a transition to standing waves, particularly in the 94 \AA\ channel. Among them, in the first three cases, a gradual transition from the sloshing phase to the standing phase is observed after a few cycles, consistent with the findings of \citet{KrishnaPrasad2021}. In contrast, in the last three cases, a standing-wave pattern is established immediately after a single reflection, resembling the behavior reported by \citet{Wang2015}. No standing phase is observed in all other cases. Since the transition depends on how quickly the higher harmonics are dissipated, such diverse behavior implies that different damping processes are at play and therefore requires a detailed theoretical modeling study to comprehend. 

The relationship between the damping time and the oscillation period also provides important insights into the damping mechanism of sloshing oscillations. This has been previously explored by \citet{Wang2005}, who analysed 51 standing slow-mode oscillation events using SUMER observations. The authors find that the relation between damping time and oscillation period for those cases can be described using a power law with an exponent of $0.96 \pm 0.18$. However, no such dependence has been explicitly investigated for sloshing oscillations so far, primarily due to the limited number of cases identified. In Figure~\ref{tau_vs_P}, we show the scaling of damping time with oscillation period separately for the AIA 131 and 94 \AA\ channels. For the 131~\AA\ channel, we find a power-law exponent of $1.38 \pm 0.12$. When data from previous studies \citep{Kumar2013, Wang2015} are included, this value becomes $1.39 \pm 0.07$. In the 94~\AA\ channel, the power-law exponent is $1.44 \pm 0.11$ for the data from this study, and when combined with the earlier studies available in this channel \citep{Kumar2015, KrishnaPrasad2021}, the exponent slightly increases to $1.56 \pm 0.10$. The Pearson correlation coefficient between the damping time and oscillation period shows a moderately positive correlation with an R value of 0.5. A similar correlation coefficient was found between the two parameters in the study by \citet{Wang2005}. However, the discrepancy in the power-law exponent values compared to their study may stem, at least in part, from the small number of events analyzed in this study, motivating a more extensive statistical investigation in the future.

\begin{figure}[!h]
    \centering
    \includegraphics[width=1\textwidth]{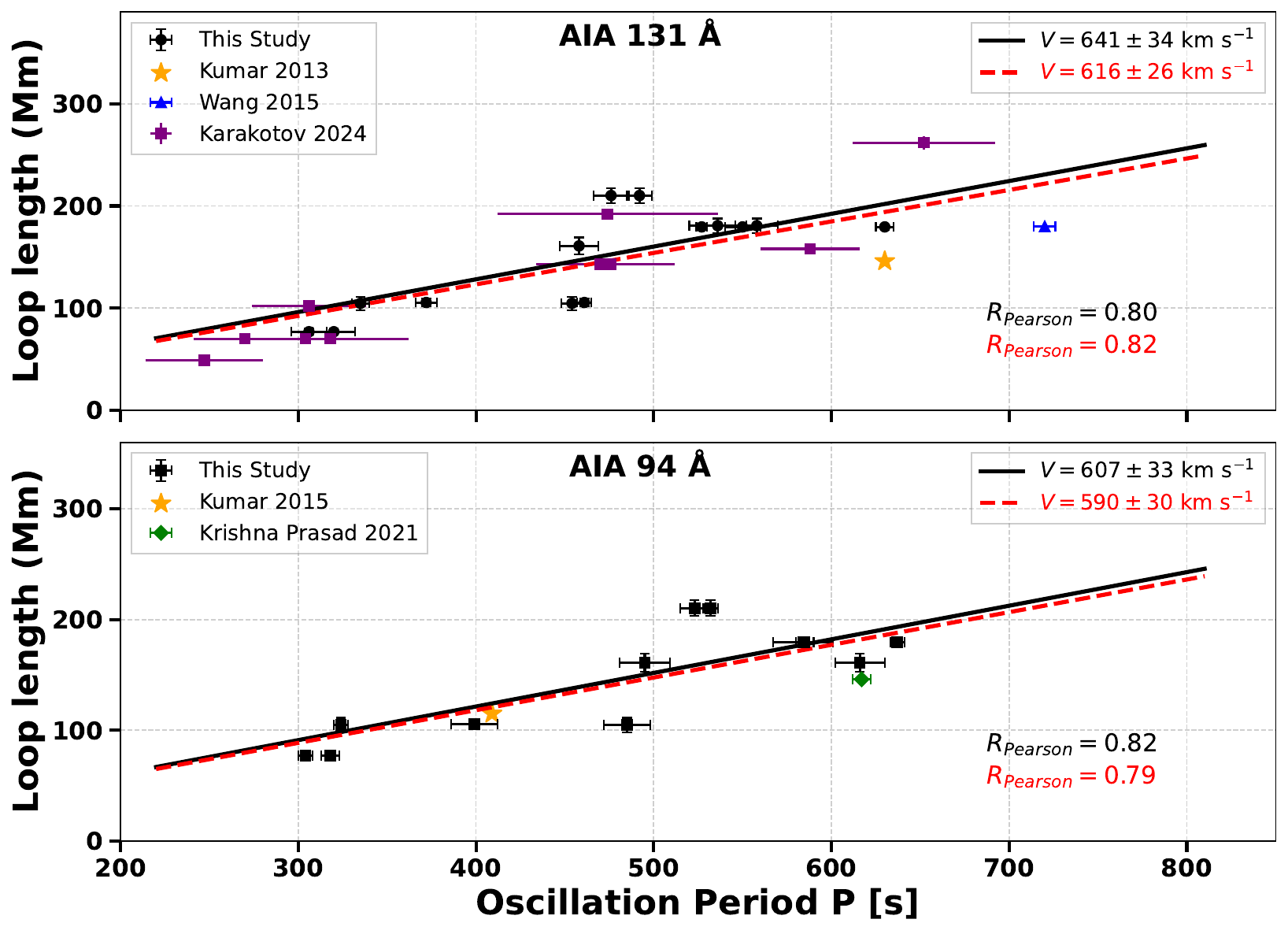}
\caption{The dependence of oscillation period on loop length in AIA 131 and 94 \AA\ channels. Black dots/squares with error bars represent results from the present study, while colored markers correspond to previous studies \citep{Kumar2013, Kumar2015, Wang2015,KrishnaPrasad2021,2024SoPh..299..163K} as denoted in the legend. The black solid line shows the best-fit linear relation for the current sample, whereas the red dashed line represents the linear fit obtained from the combined dataset. The Pearson correlation coefficients for both fits are indicated in black (current study) and red (combined study).}
    \label{vel}
\end{figure}

Another relationship that is of importance, especially in the case of standing waves, is that between the oscillation period and the loop length. The oscillation period has been shown to scale linearly with loop length, for both decayless \citep{2015A&A...583A.136A} and decaying \citep{2016A&A...585A.137G} kink oscillations, indicating a similar phase speed across different structures. \citet{2024SoPh..299..163K} have shown that a similar dependence can be found in sloshing oscillations. In Figure~\ref{vel}, we show this dependence from our data separately for the AIA 131 and 94 \AA\ channels. Additionally, we include data from all previous studies corresponding to sloshing oscillations associated with C-class flares \citep{Kumar2013, Kumar2015, Wang2015, KrishnaPrasad2021} and M-class flares \citep{2024SoPh..299..163K}. Note that from the study of \citet{2024SoPh..299..163K}, we exclude events that do not exhibit clear sloshing behavior. We also exclude the results of \citet{Mandal2016}, as that study reports only projected loop lengths, which could lead to misleading interpretations of the scaling relation. As shown in Figure~\ref{vel}, in both channels, a linear relationship between the oscillation period and loop length is observed. Moreover, the corresponding Pearson correlation coefficients listed in the plot (around 0.8) indicate a strong positive correlation between the two parameters. The slope of the fitted line, when multiplied by 2, gives a representative phase speed. This value is about 641 $\pm$ 34 \kms for the 131 \AA\ channel and 607 $\pm$ 33 \kms for the 94 \AA\ channel. When combined with the previous data from literature, these values change to 616 $\pm$ 26 \kms\, and 590 $\pm$ 30 \kms\, respectively. Note that the 94 \AA\ channel is relatively underexplored for sloshing oscillations, with only two events reported in past studies \citep{Kumar2015, KrishnaPrasad2021}. Furthermore, the analysis by \citet{KrishnaPrasad2021} did not include an independent estimation of the loop length. However, since the same loop was analyzed earlier by \citet{Kumar2013}, we adopt their value for this case. Overall, these results indicate a narrow range of phase speeds near 600 \kms applicable across different flare-impacted loop structures. We note that this value is larger than that found by \citet{2024SoPh..299..163K}. The discrepancy can again be attributed to limited data in both studies. Therefore, a larger sample will be useful to establish this behavior more conclusively.

\section{Conclusions}
\label{conclusion}

In this study, we present a unique set of observations of coronal loops exhibiting sloshing oscillations excited by successive M- and C-class flares. A set of 15 oscillation events identified within 7 distinct loops is selected and analyzed. Our analysis is restricted to the AIA 131 and 94 \AA\ channels where the oscillations are visible. All the extracted parameters are presented in Table \ref{table}. These results substantially increase the existing sample of sloshing oscillations in the literature, especially in the AIA 94 \AA\ channel. Furthermore, the unprecedented detection of sloshing oscillations triggered by both M- and C-class flares in the same loop provides a rare opportunity to study the impact of flare strength on loop and oscillation properties. As such, one of the main conclusions of our study is that oscillation properties within the same loop vary significantly across different flaring episodes. For instance, in loop 1, the damping time in the 131 \AA\ channel increases from 323$\pm$17 s during a C1.1 flare to 1012$\pm$81 s during a M2.0 flare, and decreases to 578$\pm$38 s during a C2.6 flare. The corresponding oscillation periods do not vary as drastically, but they also display a significant change. Similar changes are also observed in the 94 \AA\ channel. Because the oscillation properties depend on local physical conditions, this behavior implies changes in those conditions due to the deposition of energy by each flare. Thus, if one could decipher the damping of sloshing oscillations, such observations may open an alternative gateway to understanding the flare energetics.

In terms of oscillation properties themselves, a number of new characteristics are observed, as summarized below.

\begin{enumerate}
	\item The damping times are not always comparable to the oscillation period, as previously believed. Some events display much longer damping times, while a few others display much shorter values. This diverse behavior perhaps portrays the applicability of different damping processes across different loop structures. Additionally, the nonlinear dependence of damping time on oscillation period, as compared to the linear scaling found earlier for standing slow magnetoacoustic waves \citep{Wang2005}, highlights the gaps in our understanding of sloshing oscillations.
	\item In some cases, the cooler 94 \AA\ channel exhibits shorter damping times than the 131 \AA\ channel. This is in contradiction to the expected behavior from the thermal conduction damping \citep{KrishnaPrasad2021}. 
	\item Shorter periods ($<$ 10 min) are found even in oscillations triggered by C-class flares. While such short periods are found previously by \citet{2024SoPh..299..163K} in oscillations triggered by M-class flares, our sample expands this period range to C-class flares. However, the linear dependence of the oscillation period on loop length found may indicate that we will find more such cases in shorter loops, regardless of the flare's strength.
\end{enumerate}


\begin{acknowledgements}
AIA data are courtesy of NASA/SDO and the AIA science teams. HP acknowledges CSIR, New Delhi, for research fellowship. SKP is grateful to ISRO for a RESPOND grant (No. ISRO/RES/2/453/25-26) and to the Royal Society for the International Exchanges grant (No. IES/R3/243262). 
\end{acknowledgements}


\facilities{SDO/AIA.}

\appendix

\section{Loop Length Estimation}
\label{app}
\begin{figure}[!h]
    \centering
    \includegraphics[width=1\textwidth]{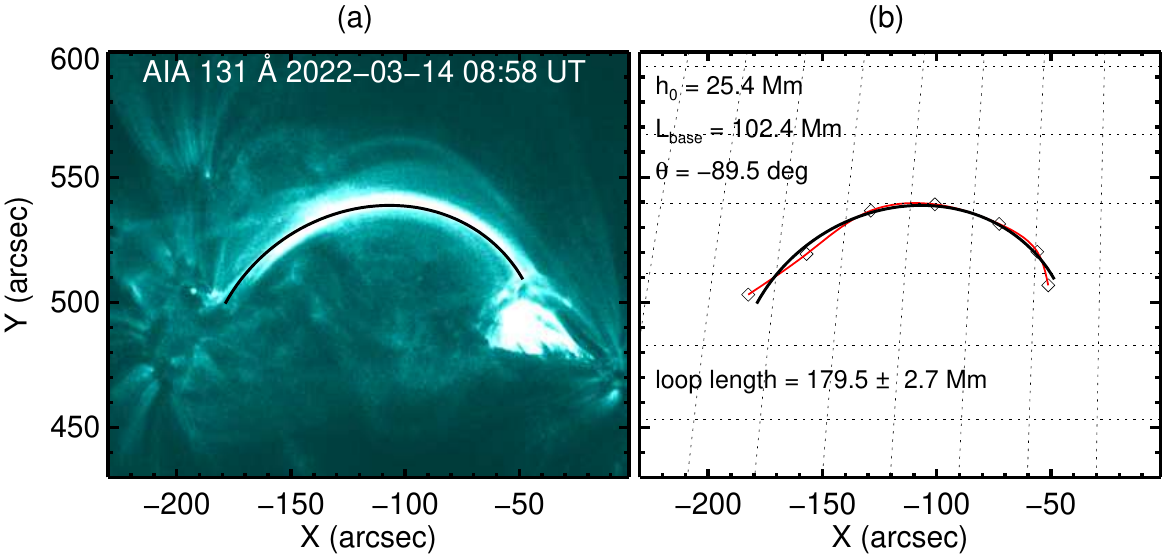}
   \caption{Estimation of loop length following the model of \cite{2002SoPh..206...99A}. (a) Snapshot of loop 1 in AIA 131~\AA\ channel selected from an instant when the entire loop structure is clearly visible. The solid black curve shows the best fit obtained from the model. (b) Diamond markers indicate the locations of manually selected points along the loop for fitting the model. The geometry of the loop is highlighted by the red curve, while the fit is shown as a black curve. Various parameters obtained, including the loop length, are listed in the plot.}
    \label{loop}
\end{figure}

The length of each coronal loop in our sample is estimated following the best-fit circular loop model of \citet{2002SoPh..206...99A} as demonstrated in Figure \ref{loop} for loop 1. The loop length is calculated as \(L=\pi r\), where the loop radius \(r\) is determined from the height of the loop centre, \(h_0\), and the baseline length between the two footpoints, \(L_{\rm base}\), as

\begin{equation}
r = \sqrt{h_0^2 + \left(\frac{L_{\rm base}}{2}\right)^2 }
\end{equation}

\bibliography{references}{}

@ARTICLE{2016A&A...585A.137G,
       author = {{Goddard}, C.~R. and {Nistic{\`o}}, G. and {Nakariakov}, V.~M. and {Zimovets}, I.~V.},
        title = "{A statistical study of decaying kink oscillations detected using SDO/AIA}",
      journal = {\aap},
         year = 2016,
        month = jan,
       volume = {585},
          eid = {A137},
        pages = {A137},
          doi = {10.1051/0004-6361/201527341},
archivePrefix = {arXiv},
       eprint = {1511.03558},
 primaryClass = {astro-ph.SR},
       adsurl = {https://ui.adsabs.harvard.edu/abs/2016A&A...585A.137G}
}

@ARTICLE{2015A&A...583A.136A,
       author = {{Anfinogentov}, S.~A. and {Nakariakov}, V.~M. and {Nistic{\`o}}, G.},
        title = "{Decayless low-amplitude kink oscillations: a common phenomenon in the solar corona?}",
      journal = {\aap},
         year = 2015,
        month = nov,
       volume = {583},
          eid = {A136},
        pages = {A136},
          doi = {10.1051/0004-6361/201526195},
archivePrefix = {arXiv},
       eprint = {1509.05519},
 primaryClass = {astro-ph.SR},
       adsurl = {https://ui.adsabs.harvard.edu/abs/2015A&A...583A.136A}
}

@ARTICLE{Banerjee2021,
       author = {{Banerjee}, D. and {Krishna Prasad}, S. and {Pant}, V. and {McLaughlin}, J.~A. and {Antolin}, P. and {Magyar}, N. and {Ofman}, L. and {Tian}, H. and {Van Doorsselaere}, T. and {De Moortel}, I. and {Wang}, T.~J.},
        title = "{Magnetohydrodynamic Waves in Open Coronal Structures}",
      journal = {\ssr},
         year = 2021,
        month = oct,
       volume = {217},
       number = {7},
          eid = {76},
        pages = {76},
          doi = {10.1007/s11214-021-00849-0},
archivePrefix = {arXiv},
       eprint = {2012.08802},
 primaryClass = {astro-ph.SR},
       adsurl = {https://ui.adsabs.harvard.edu/abs/2021SSRv..217...76B}
}

@ARTICLE{XRS/GOES-R,
       author = {{Woods}, Thomas N. and {Eden}, Thomas and {Eparvier}, Francis G. and {Jones}, Andrew R. and {Woodraska}, Donald L. and {Chamberlin}, Phillip C. and {Machol}, Janet L.},
        title = "{GOES-R Series X-Ray Sensor (XRS): 1. Design and Pre-Flight Calibration}",
      journal = {Journal of Geophysical Research (Space Physics)},
         year = 2024,
        month = nov,
       volume = {129},
       number = {11},
        pages = {2024JA032925},
          doi = {10.1029/2024JA032925},
       adsurl = {https://ui.adsabs.harvard.edu/abs/2024JGRA..12932925W}
}

@INPROCEEDINGS{GOES,
       author = {{Hanser}, Frederick A. and {Sellers}, Francis B.},
        title = "{Design and calibration of the GOES-8 solar x-ray sensor: the XRS}",
    booktitle = {GOES-8 and Beyond},
         year = 1996,
       editor = {{Washwell}, Edward R.},
       series = {Society of Photo-Optical Instrumentation Engineers (SPIE) Conference Series},
       volume = {2812},
        month = oct,
        pages = {344-352},
          doi = {10.1117/12.254082},
       adsurl = {https://ui.adsabs.harvard.edu/abs/1996SPIE.2812..344H}
}

@ARTICLE{AIA,
       author = {{Lemen}, James R. and {Title}, Alan M. and {Akin}, David J. and {Boerner}, Paul F. and {Chou}, Catherine and {Drake}, Jerry F. and {Duncan}, Dexter W. and {Edwards}, Christopher G. and {Friedlaender}, Frank M. and {Heyman}, Gary F. and {Hurlburt}, Neal E. and {Katz}, Noah L. and {Kushner}, Gary D. and {Levay}, Michael and {Lindgren}, Russell W. and {Mathur}, Dnyanesh P. and {McFeaters}, Edward L. and {Mitchell}, Sarah and {Rehse}, Roger A. and {Schrijver}, Carolus J. and {Springer}, Larry A. and {Stern}, Robert A. and {Tarbell}, Theodore D. and {Wuelser}, Jean-Pierre and {Wolfson}, C. Jacob and {Yanari}, Carl and {Bookbinder}, Jay A. and {Cheimets}, Peter N. and {Caldwell}, David and {Deluca}, Edward E. and {Gates}, Richard and {Golub}, Leon and {Park}, Sang and {Podgorski}, William A. and {Bush}, Rock I. and {Scherrer}, Philip H. and {Gummin}, Mark A. and {Smith}, Peter and {Auker}, Gary and {Jerram}, Paul and {Pool}, Peter and {Soufli}, Regina and {Windt}, David L. and {Beardsley}, Sarah and {Clapp}, Matthew and {Lang}, James and {Waltham}, Nicholas},
        title = "{The Atmospheric Imaging Assembly (AIA) on the Solar Dynamics Observatory (SDO)}",
      journal = {\solphys},
         year = 2012,
        month = jan,
       volume = {275},
       number = {1-2},
        pages = {17-40},
          doi = {10.1007/s11207-011-9776-8},
       adsurl = {https://ui.adsabs.harvard.edu/abs/2012SoPh..275...17L}
}

@ARTICLE{SDO,
       author = {{Pesnell}, W. Dean and {Thompson}, B.~J. and {Chamberlin}, P.~C.},
        title = "{The Solar Dynamics Observatory (SDO)}",
      journal = {\solphys},
         year = 2012,
        month = jan,
       volume = {275},
       number = {1-2},
        pages = {3-15},
          doi = {10.1007/s11207-011-9841-3},
       adsurl = {https://ui.adsabs.harvard.edu/abs/2012SoPh..275....3P}
}

@ARTICLE{Wang2003,
       author = {{Wang}, T.~J. and {Solanki}, S.~K. and {Curdt}, W. and {Innes}, D.~E. and {Dammasch}, I.~E. and {Kliem}, B.},
        title = "{Hot coronal loop oscillations observed with SUMER: Examples and statistics}",
      journal = {\aap},
         year = 2003,
        month = aug,
       volume = {406},
        pages = {1105-1121},
          doi = {10.1051/0004-6361:20030858},
       adsurl = {https://ui.adsabs.harvard.edu/abs/2003A&A...406.1105W}
}

@ARTICLE{Wang2002,
       author = {{Wang}, Tongjiang and {Solanki}, S.~K. and {Curdt}, W. and {Innes}, D.~E. and {Dammasch}, I.~E.},
        title = "{Doppler Shift Oscillations of Hot Solar Coronal Plasma Seen by SUMER: A Signature of Loop Oscillations?}",
      journal = {\apjl},
         year = 2002,
        month = jul,
       volume = {574},
       number = {1},
        pages = {L101-L104},
          doi = {10.1086/342189},
       adsurl = {https://ui.adsabs.harvard.edu/abs/2002ApJ...574L.101W}
}

@ARTICLE{Wang2005,
       author = {{Wang}, T.~J. and {Solanki}, S.~K. and {Innes}, D.~E. and {Curdt}, W.},
        title = "{Initiation of hot coronal loop oscillations: Spectral features}",
      journal = {\aap},
         year = 2005,
        month = may,
       volume = {435},
       number = {2},
        pages = {753-764},
          doi = {10.1051/0004-6361:20052680},
       adsurl = {https://ui.adsabs.harvard.edu/abs/2005A&A...435..753W}
}

@ARTICLE{Kumar2013,
       author = {{Kumar}, Pankaj and {Innes}, D.~E. and {Inhester}, B.},
        title = "{Solar Dynamics Observatory/Atmospheric Imaging Assembly Observations of a Reflecting Longitudinal Wave in a Coronal Loop}",
      journal = {\apjl},
         year = 2013,
        month = dec,
       volume = {779},
       number = {1},
          eid = {L7},
        pages = {L7},
          doi = {10.1088/2041-8205/779/1/L7},
archivePrefix = {arXiv},
       eprint = {1409.3896},
 primaryClass = {astro-ph.SR},
       adsurl = {https://ui.adsabs.harvard.edu/abs/2013ApJ...779L...7K}
}

@ARTICLE{Kumar2015,
       author = {{Kumar}, Pankaj and {Nakariakov}, Valery M. and {Cho}, Kyung-Suk},
        title = "{X-Ray and EUV Observations of Simultaneous Short and Long Period Oscillations in Hot Coronal Arcade Loops}",
      journal = {\apj},
         year = 2015,
        month = may,
       volume = {804},
       number = {1},
          eid = {4},
        pages = {4},
          doi = {10.1088/0004-637X/804/1/4},
archivePrefix = {arXiv},
       eprint = {1502.07117},
 primaryClass = {astro-ph.SR},
       adsurl = {https://ui.adsabs.harvard.edu/abs/2015ApJ...804....4K}
}

@ARTICLE{KrishnaPrasad2021,
       author = {{Krishna Prasad}, S. and {Van Doorsselaere}, T.},
        title = "{Compressive Oscillations in Hot Coronal Loops: Are Sloshing Oscillations and Standing Slow Waves Independent?}",
      journal = {\apj},
         year = 2021,
        month = jun,
       volume = {914},
       number = {2},
          eid = {81},
        pages = {81},
          doi = {10.3847/1538-4357/abfb01},
archivePrefix = {arXiv},
       eprint = {2104.12038},
 primaryClass = {astro-ph.SR},
       adsurl = {https://ui.adsabs.harvard.edu/abs/2021ApJ...914...81K}
}

@ARTICLE{Ofman2002,
       author = {{Ofman}, L. and {Wang}, Tongjiang},
        title = "{Hot Coronal Loop Oscillations Observed by SUMER: Slow Magnetosonic Wave Damping by Thermal Conduction}",
      journal = {\apjl},
         year = 2002,
        month = nov,
       volume = {580},
       number = {1},
        pages = {L85-L88},
          doi = {10.1086/345548},
       adsurl = {https://ui.adsabs.harvard.edu/abs/2002ApJ...580L..85O}
}

@ARTICLE{Wang2015,
       author = {{Wang}, Tongjiang and {Ofman}, Leon and {Sun}, Xudong and {Provornikova}, Elena and {Davila}, Joseph M.},
        title = "{Evidence of Thermal Conduction Suppression in a Solar Flaring Loop by Coronal Seismology of Slow-mode Waves}",
      journal = {\apjl},
         year = 2015,
        month = sep,
       volume = {811},
       number = {1},
          eid = {L13},
        pages = {L13},
          doi = {10.1088/2041-8205/811/1/L13},
archivePrefix = {arXiv},
       eprint = {1509.00920},
 primaryClass = {astro-ph.SR},
       adsurl = {https://ui.adsabs.harvard.edu/abs/2015ApJ...811L..13W}
}

@ARTICLE{Wang2018,
       author = {{Wang}, Tongjiang and {Ofman}, Leon and {Sun}, Xudong and {Solanki}, Sami K. and {Davila}, Joseph M.},
        title = "{Effect of Transport Coefficients on Excitation of Flare-induced Standing Slow-mode Waves in Coronal Loops}",
      journal = {\apj},
         year = 2018,
        month = jun,
       volume = {860},
       number = {2},
          eid = {107},
        pages = {107},
          doi = {10.3847/1538-4357/aac38a},
archivePrefix = {arXiv},
       eprint = {1805.03282},
 primaryClass = {astro-ph.SR},
       adsurl = {https://ui.adsabs.harvard.edu/abs/2018ApJ...860..107W}
}

@ARTICLE{Nakariakov2017,
       author = {{Nakariakov}, V.~M. and {Afanasyev}, A.~N. and {Kumar}, S. and {Moon}, Y. -J.},
        title = "{Effect of Local Thermal Equilibrium Misbalance on Long-wavelength Slow Magnetoacoustic Waves}",
      journal = {\apj},
         year = 2017,
        month = nov,
       volume = {849},
       number = {1},
          eid = {62},
        pages = {62},
          doi = {10.3847/1538-4357/aa8ea3},
       adsurl = {https://ui.adsabs.harvard.edu/abs/2017ApJ...849...62N}
}

@ARTICLE{Reale2016,
       author = {{Reale}, F.},
        title = "{Plasma Sloshing in Pulse-heated Solar and Stellar Coronal Loops}",
      journal = {\apjl},
         year = 2016,
        month = aug,
       volume = {826},
       number = {2},
          eid = {L20},
        pages = {L20},
          doi = {10.3847/2041-8205/826/2/L20},
archivePrefix = {arXiv},
       eprint = {1607.01329},
 primaryClass = {astro-ph.SR},
       adsurl = {https://ui.adsabs.harvard.edu/abs/2016ApJ...826L..20R}
}

@ARTICLE{Mandal2016,
       author = {{Mandal}, Sudip and {Yuan}, Ding and {Fang}, Xia and {Banerjee}, Dipankar and {Pant}, Vaibhav and {Van Doorsselaere}, Tom},
        title = "{Reflection of Propagating Slow Magneto-acoustic Waves in Hot Coronal Loops: Multi-instrument Observations and Numerical Modeling}",
      journal = {\apj},
         year = 2016,
        month = sep,
       volume = {828},
       number = {2},
          eid = {72},
        pages = {72},
          doi = {10.3847/0004-637X/828/2/72},
archivePrefix = {arXiv},
       eprint = {1604.08133},
 primaryClass = {astro-ph.SR},
       adsurl = {https://ui.adsabs.harvard.edu/abs/2016ApJ...828...72M}
}

@ARTICLE{Yuan2012,
       author = {{Yuan}, D. and {Nakariakov}, V.~M.},
        title = "{Measuring the apparent phase speed of propagating EUV disturbances}",
      journal = {\aap},
         year = 2012,
        month = jul,
       volume = {543},
          eid = {A9},
        pages = {A9},
          doi = {10.1051/0004-6361/201218848},
       adsurl = {https://ui.adsabs.harvard.edu/abs/2012A&A...543A...9Y}
}

@ARTICLE{2002SoPh..206...99A,
       author = {{Aschwanden}, Markus J. and {De Pontieu}, Bart and {Schrijver}, Carolus J. and {Title}, Alan M.},
        title = "{Transverse Oscillations in Coronal Loops Observed with TRACE   II. Measurements of Geometric and Physical Parameters}",
      journal = {\solphys},
         year = 2002,
        month = mar,
       volume = {206},
       number = {1},
        pages = {99-132},
          doi = {10.1023/A:1014916701283},
       adsurl = {https://ui.adsabs.harvard.edu/abs/2002SoPh..206...99A}
}

@ARTICLE{2024SoPh..299..163K,
       author = {{Karakotov}, Ruslan and {Kuznetsov}, Alexey and {Anfinogentov}, Sergey and {Nakariakov}, Valery M.},
        title = "{Sloshing Oscillations in Hot Coronal Loops Associated with M-Class Flares}",
      journal = {\solphys},
         year = 2024,
        month = dec,
       volume = {299},
       number = {12},
          eid = {163},
        pages = {163},
          doi = {10.1007/s11207-024-02404-w},
       adsurl = {https://ui.adsabs.harvard.edu/abs/2024SoPh..299..163K}
}

@ARTICLE{2012A&A...539A.146H,
       author = {{Hannah}, I.~G. and {Kontar}, E.~P.},
        title = "{Differential emission measures from the regularized inversion of Hinode and SDO data}",
      journal = {\aap},
         year = 2012,
        month = mar,
       volume = {539},
          eid = {A146},
        pages = {A146},
          doi = {10.1051/0004-6361/201117576},
archivePrefix = {arXiv},
       eprint = {1201.2642},
 primaryClass = {astro-ph.SR},
       adsurl = {https://ui.adsabs.harvard.edu/abs/2012A&A...539A.146H}
}

@ARTICLE{2014SoPh..289.3233L,
       author = {{Liu}, Wei and {Ofman}, Leon},
        title = "{Advances in Observing Various Coronal EUV Waves in the SDO Era and Their Seismological Applications (Invited Review)}",
      journal = {\solphys},
         year = 2014,
        month = sep,
       volume = {289},
       number = {9},
        pages = {3233-3277},
          doi = {10.1007/s11207-014-0528-4},
archivePrefix = {arXiv},
       eprint = {1404.0670},
 primaryClass = {astro-ph.SR},
       adsurl = {https://ui.adsabs.harvard.edu/abs/2014SoPh..289.3233L}
}

@ARTICLE{2012RSPTA.370.3193D,
       author = {{De Moortel}, I. and {Nakariakov}, V.~M.},
        title = "{Magnetohydrodynamic waves and coronal seismology: an overview of recent results}",
      journal = {Philosophical Transactions of the Royal Society of London Series A},
         year = 2012,
        month = jul,
       volume = {370},
       number = {1970},
        pages = {3193-3216},
          doi = {10.1098/rsta.2011.0640},
archivePrefix = {arXiv},
       eprint = {1202.1944},
 primaryClass = {astro-ph.SR},
       adsurl = {https://ui.adsabs.harvard.edu/abs/2012RSPTA.370.3193D}
}

@ARTICLE{2000SoPh..193..139R,
       author = {{Roberts}, B.},
        title = "{Waves and Oscillations in the Corona - (Invited Review)}",
      journal = {\solphys},
         year = 2000,
        month = apr,
       volume = {193},
        pages = {139-152},
          doi = {10.1023/A:1005237109398},
       adsurl = {https://ui.adsabs.harvard.edu/abs/2000SoPh..193..139R}
}

@ARTICLE{2018AdSpR..61..655A,
       author = {{Arregui}, I{\~n}igo},
        title = "{Bayesian coronal seismology}",
      journal = {Advances in Space Research},
         year = 2018,
        month = jan,
       volume = {61},
       number = {2},
        pages = {655-672},
          doi = {10.1016/j.asr.2017.09.031},
archivePrefix = {arXiv},
       eprint = {1709.08372},
 primaryClass = {astro-ph.SR},
       adsurl = {https://ui.adsabs.harvard.edu/abs/2018AdSpR..61..655A}
}

@ARTICLE{2020SSRv..216..140V,
       author = {{Van Doorsselaere}, Tom and {Srivastava}, Abhishek K. and {Antolin}, Patrick and {Magyar}, Norbert and {Vasheghani Farahani}, Soheil and {Tian}, Hui and {Kolotkov}, Dmitrii and {Ofman}, Leon and {Guo}, Mingzhe and {Arregui}, I{\~n}igo and {De Moortel}, Ineke and {Pascoe}, David},
        title = "{Coronal Heating by MHD Waves}",
      journal = {\ssr},
         year = 2020,
        month = dec,
       volume = {216},
       number = {8},
          eid = {140},
        pages = {140},
          doi = {10.1007/s11214-020-00770-y},
archivePrefix = {arXiv},
       eprint = {2012.01371},
 primaryClass = {astro-ph.SR},
       adsurl = {https://ui.adsabs.harvard.edu/abs/2020SSRv..216..140V}
}

@ARTICLE{2021SSRv..217...34W,
       author = {{Wang}, Tongjiang and {Ofman}, Leon and {Yuan}, Ding and {Reale}, Fabio and {Kolotkov}, Dmitrii Y. and {Srivastava}, Abhishek K.},
        title = "{Slow-Mode Magnetoacoustic Waves in Coronal Loops}",
      journal = {\ssr},
         year = 2021,
        month = mar,
       volume = {217},
       number = {2},
          eid = {34},
        pages = {34},
          doi = {10.1007/s11214-021-00811-0},
archivePrefix = {arXiv},
       eprint = {2102.11376},
 primaryClass = {astro-ph.SR},
       adsurl = {https://ui.adsabs.harvard.edu/abs/2021SSRv..217...34W}
}

@ARTICLE{2003A&A...402L..17W,
       author = {{Wang}, T.~J. and {Solanki}, S.~K. and {Innes}, D.~E. and {Curdt}, W. and {Marsch}, E.},
        title = "{Slow-mode standing waves observed by SUMER  in hot coronal loops}",
      journal = {\aap},
         year = 2003,
        month = may,
       volume = {402},
        pages = {L17-L20},
          doi = {10.1051/0004-6361:20030448},
       adsurl = {https://ui.adsabs.harvard.edu/abs/2003A&A...402L..17W}
}

@ARTICLE{2005ApJ...620L..67M,
       author = {{Mariska}, John T.},
        title = "{Observations of Solar Flare Doppler Shift Oscillations with the Bragg Crystal Spectrometer on Yohkoh}",
      journal = {\apjl},
         year = 2005,
        month = feb,
       volume = {620},
       number = {1},
        pages = {L67-L70},
          doi = {10.1086/428611},
archivePrefix = {arXiv},
       eprint = {astro-ph/0501093},
 primaryClass = {astro-ph},
       adsurl = {https://ui.adsabs.harvard.edu/abs/2005ApJ...620L..67M}
}

@ARTICLE{2006ApJ...639..484M,
       author = {{Mariska}, John T.},
        title = "{Characteristics of Solar Flare Doppler-Shift Oscillations Observed with the Bragg Crystal Spectrometer on Yohkoh}",
      journal = {\apj},
         year = 2006,
        month = mar,
       volume = {639},
       number = {1},
        pages = {484-494},
          doi = {10.1086/499296},
archivePrefix = {arXiv},
       eprint = {astro-ph/0511070},
 primaryClass = {astro-ph},
       adsurl = {https://ui.adsabs.harvard.edu/abs/2006ApJ...639..484M}
}

@ARTICLE{2008ApJ...681L..41M,
       author = {{Mariska}, John T. and {Warren}, Harry P. and {Williams}, David R. and {Watanabe}, Tetsuya},
        title = "{Observations of Doppler Shift Oscillations with the EUV Imaging Spectrometer on Hinode}",
      journal = {\apjl},
         year = 2008,
        month = jul,
       volume = {681},
       number = {1},
        pages = {L41},
          doi = {10.1086/590341},
archivePrefix = {arXiv},
       eprint = {0806.0265},
 primaryClass = {astro-ph},
       adsurl = {https://ui.adsabs.harvard.edu/abs/2008ApJ...681L..41M}
}

@ARTICLE{2019A&A...628A.133K,
       author = {{Kolotkov}, D.~Y. and {Nakariakov}, V.~M. and {Zavershinskii}, D.~I.},
        title = "{Damping of slow magnetoacoustic oscillations by the misbalance between heating and cooling processes in the solar corona}",
      journal = {\aap},
         year = 2019,
        month = aug,
       volume = {628},
          eid = {A133},
        pages = {A133},
          doi = {10.1051/0004-6361/201936072},
archivePrefix = {arXiv},
       eprint = {1907.07051},
 primaryClass = {astro-ph.SR},
       adsurl = {https://ui.adsabs.harvard.edu/abs/2019A&A...628A.133K}
}

@ARTICLE{2005A&A...436..701S,
       author = {{Selwa}, M. and {Murawski}, K. and {Solanki}, S.~K.},
        title = "{Excitation and damping of slow magnetosonic standing waves in a solar coronal loop}",
      journal = {\aap},
         year = 2005,
        month = jun,
       volume = {436},
       number = {2},
        pages = {701-709},
          doi = {10.1051/0004-6361:20042319},
       adsurl = {https://ui.adsabs.harvard.edu/abs/2005A&A...436..701S}
}

@ARTICLE{2009AnGeo..27.3899S,
       author = {{Selwa}, M. and {Ofman}, L.},
        title = "{3-D numerical simulations of coronal loops oscillations}",
      journal = {Annales Geophysicae},
         year = 2009,
        month = oct,
       volume = {27},
       number = {10},
        pages = {3899-3908},
          doi = {10.5194/angeo-27-3899-2009},
       adsurl = {https://ui.adsabs.harvard.edu/abs/2009AnGeo..27.3899S}
}

@ARTICLE{2012ApJ...754..111O,
       author = {{Ofman}, L. and {Wang}, T.~J. and {Davila}, J.~M.},
        title = "{Slow Magnetosonic Waves and Fast Flows in Active Region Loops}",
      journal = {\apj},
         year = 2012,
        month = aug,
       volume = {754},
       number = {2},
          eid = {111},
        pages = {111},
          doi = {10.1088/0004-637X/754/2/111},
archivePrefix = {arXiv},
       eprint = {1205.5732},
 primaryClass = {astro-ph.SR},
       adsurl = {https://ui.adsabs.harvard.edu/abs/2012ApJ...754..111O}
}

@ARTICLE{2008A&A...483..301B,
       author = {{Bradshaw}, S.~J. and {Erd{\'e}lyi}, R.},
        title = "{Radiative damping of standing acoustic waves in solar coronal loops}",
      journal = {\aap},
         year = 2008,
        month = may,
       volume = {483},
       number = {1},
        pages = {301-309},
          doi = {10.1051/0004-6361:20079128},
       adsurl = {https://ui.adsabs.harvard.edu/abs/2008A&A...483..301B}
}

@ARTICLE{2008ApJ...685.1286V,
       author = {{Verwichte}, E. and {Haynes}, M. and {Arber}, T.~D. and {Brady}, C.~S.},
        title = "{Damping of Slow MHD Coronal Loop Oscillations by Shocks}",
      journal = {\apj},
         year = 2008,
        month = oct,
       volume = {685},
       number = {2},
        pages = {1286-1290},
          doi = {10.1086/591077},
       adsurl = {https://ui.adsabs.harvard.edu/abs/2008ApJ...685.1286V}
}

@ARTICLE{2013A&A...553A..23R,
       author = {{Ruderman}, M.~S.},
        title = "{Nonlinear damped standing slow waves in hot coronal magnetic loops}",
      journal = {\aap},
         year = 2013,
        month = may,
       volume = {553},
          eid = {A23},
        pages = {A23},
          doi = {10.1051/0004-6361/201321175},
       adsurl = {https://ui.adsabs.harvard.edu/abs/2013A&A...553A..23R}
}

@ARTICLE{2014ApJ...786...36A,
       author = {{Al-Ghafri}, K.~S. and {Ruderman}, M.~S. and {Williamson}, A. and {Erd{\'e}lyi}, R.},
        title = "{Longitudinal Magnetohydrodynamics Oscillations in Dissipative, Cooling Coronal Loops}",
      journal = {\apj},
         year = 2014,
        month = may,
       volume = {786},
       number = {1},
          eid = {36},
        pages = {36},
          doi = {10.1088/0004-637X/786/1/36},
       adsurl = {https://ui.adsabs.harvard.edu/abs/2014ApJ...786...36A}
}

@ARTICLE{2004A&A...414L..25N,
       author = {{Nakariakov}, V.~M. and {Tsiklauri}, D. and {Kelly}, A. and {Arber}, T.~D. and {Aschwanden}, M.~J.},
        title = "{Acoustic oscillations in solar and stellar flaring loops}",
      journal = {\aap},
         year = 2004,
        month = jan,
       volume = {414},
        pages = {L25-L28},
          doi = {10.1051/0004-6361:20031738},
archivePrefix = {arXiv},
       eprint = {astro-ph/0402223},
 primaryClass = {astro-ph},
       adsurl = {https://ui.adsabs.harvard.edu/abs/2004A&A...414L..25N}
}

@ARTICLE{2006A&A...446.1151N,
       author = {{Nakariakov}, V.~M. and {Melnikov}, V.~F.},
        title = "{Modulation of gyrosynchrotron emission in solar and stellar flares by slow magnetoacoustic oscillations}",
      journal = {\aap},
         year = 2006,
        month = feb,
       volume = {446},
       number = {3},
        pages = {1151-1156},
          doi = {10.1051/0004-6361:20053944},
       adsurl = {https://ui.adsabs.harvard.edu/abs/2006A&A...446.1151N}
}

@ARTICLE{2015ApJ...807...98Y,
       author = {{Yuan}, D. and {Van Doorsselaere}, T. and {Banerjee}, D. and {Antolin}, P.},
        title = "{Forward Modeling of Standing Slow Modes in Flaring Coronal Loops}",
      journal = {\apj},
         year = 2015,
        month = jul,
       volume = {807},
       number = {1},
          eid = {98},
        pages = {98},
          doi = {10.1088/0004-637X/807/1/98},
archivePrefix = {arXiv},
       eprint = {1504.07475},
 primaryClass = {astro-ph.SR},
       adsurl = {https://ui.adsabs.harvard.edu/abs/2015ApJ...807...98Y}
}

@ARTICLE{2015ApJ...813...33F,
       author = {{Fang}, X. and {Yuan}, D. and {Van Doorsselaere}, T. and {Keppens}, R. and {Xia}, C.},
        title = "{Modeling of Reflective Propagating Slow-mode Wave in a Flaring Loop}",
      journal = {\apj},
         year = 2015,
        month = nov,
       volume = {813},
       number = {1},
          eid = {33},
        pages = {33},
          doi = {10.1088/0004-637X/813/1/33},
archivePrefix = {arXiv},
       eprint = {1509.04536},
 primaryClass = {astro-ph.SR},
       adsurl = {https://ui.adsabs.harvard.edu/abs/2015ApJ...813...33F}
}

@ARTICLE{2012ApJ...756L..36K,
       author = {{Kim}, S. and {Nakariakov}, V.~M. and {Shibasaki}, K.},
        title = "{Slow Magnetoacoustic Oscillations in the Microwave Emission of Solar Flares}",
      journal = {\apjl},
         year = 2012,
        month = sep,
       volume = {756},
       number = {2},
          eid = {L36},
        pages = {L36},
          doi = {10.1088/2041-8205/756/2/L36},
archivePrefix = {arXiv},
       eprint = {1310.2796},
 primaryClass = {astro-ph.SR},
       adsurl = {https://ui.adsabs.harvard.edu/abs/2012ApJ...756L..36K}
}

@ARTICLE{Berghmans2001,
       author = {{Berghmans}, D. and {McKenzie}, D. and {Clette}, F.},
        title = "{Active region transient brightenings. A simultaneous view by SXT, EIT and TRACE}",
      journal = {\aap},
         year = 2001,
        month = apr,
       volume = {369},
        pages = {291-304},
          doi = {10.1051/0004-6361:20010142},
       adsurl = {https://ui.adsabs.harvard.edu/abs/2001A&A...369..291B}
}

\section{Time-Distance maps and extraction of wave properties for loops ~2--7}
\label{events}
This appendix presents the time--distance maps and the corresponding extraction of oscillation parameters (period and damping time) for loops~2--7 in Table~\ref{table}, following the same procedure described in Figures~\ref{xt} and \ref{fitting}. In the case of Event \textbf{a} in loop 5 (see Fig. \ref{fig:fig12}), only a single bright ridge is visible in the 131~\AA\ time--distance map, preventing a reliable determination of the oscillation parameters. All other events exhibit clear oscillatory signatures, allowing the wave properties to be measured.
\begin{figure}[!h]
    \centering
    \includegraphics[width=.7\textwidth]{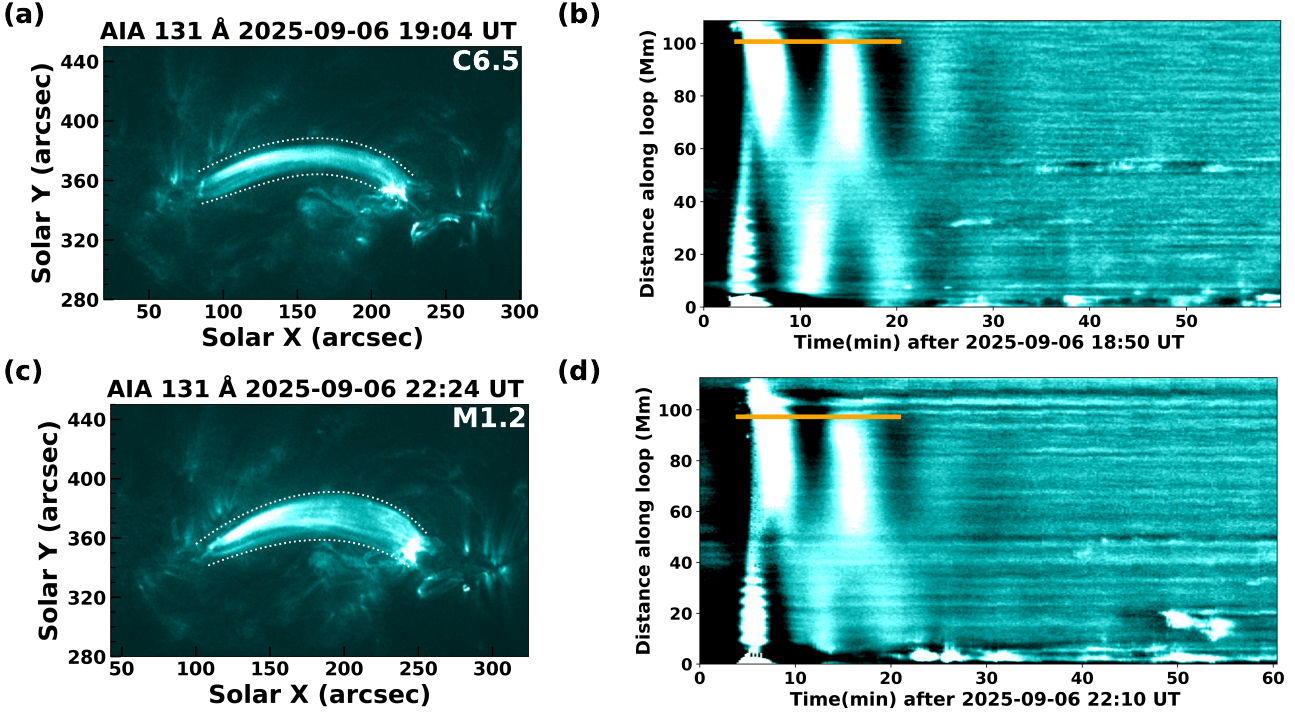}
   \caption{Similar to figure~\ref{xt} but for Loop~2 in table~\ref{table}.}
\end{figure}

\begin{figure}[!h]
    \centering
    \includegraphics[width=.6\textwidth]{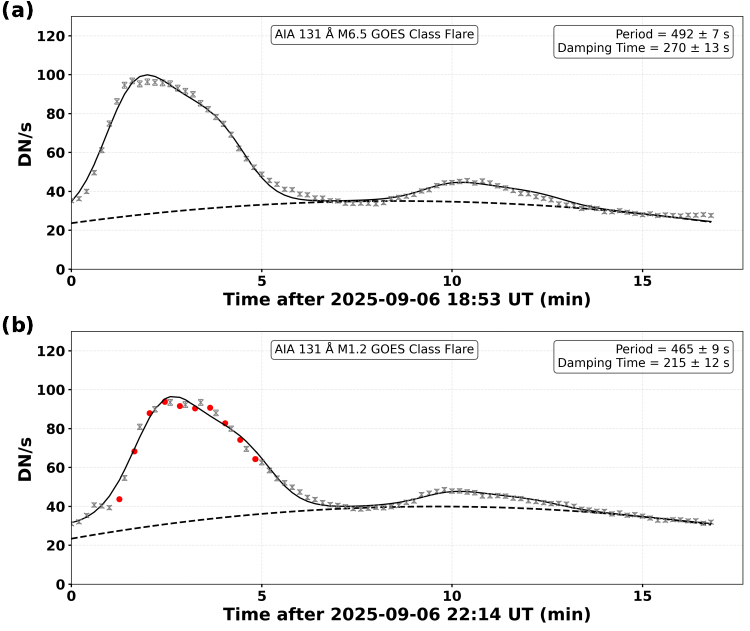}
   \caption{Similar to figure~\ref{fitting} but for Loop~2 in table~\ref{table}. The data points shown by red markers in the bottom panel correspond to frames with low exposure times and were excluded from the fitting procedure.}
\end{figure}

\begin{figure}[!h]
    \centering
    \includegraphics[width=.8\textwidth]{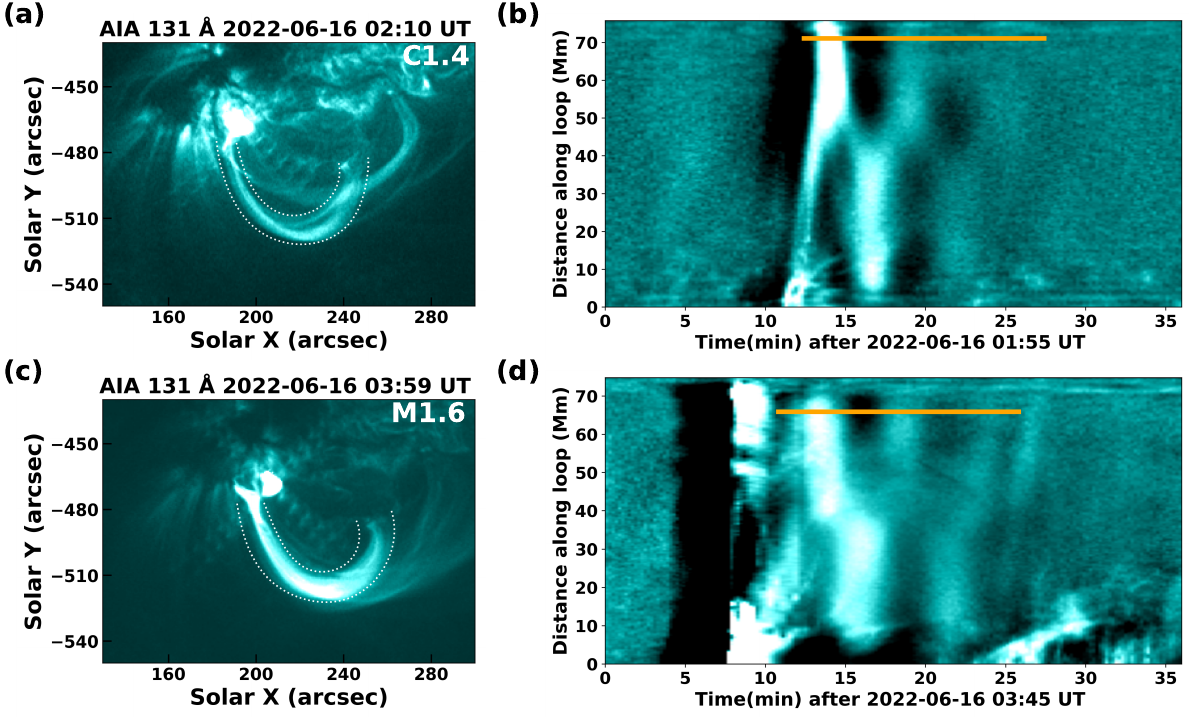}
   \caption{Similar to figure~\ref{xt} but for Loop~3 in table~\ref{table}.}
\end{figure}

\begin{figure}[!h]
    \centering
    \includegraphics[width=.7\textwidth]{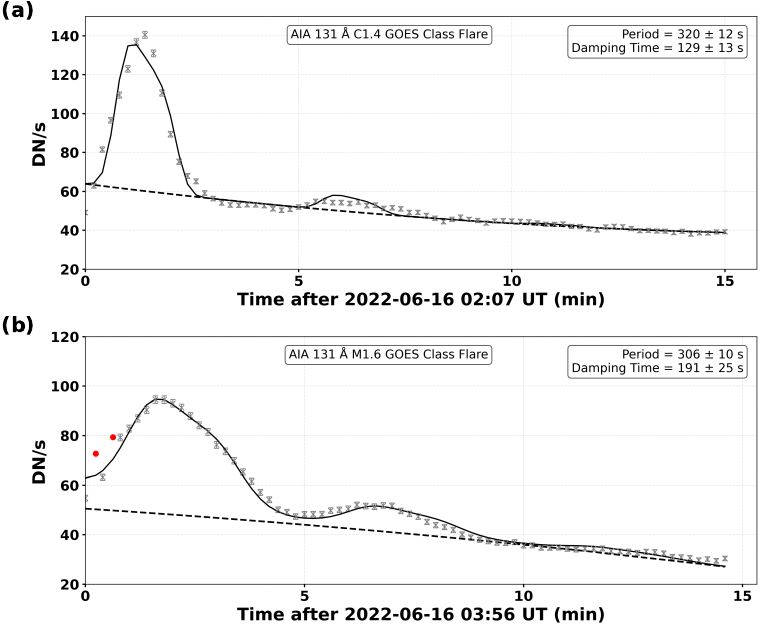}
   \caption{Similar to figure~\ref{fitting} but for Loop~3 in table~\ref{table}.}
\end{figure}

\begin{figure}[!h]
    \centering
    \includegraphics[width=.7\textwidth]{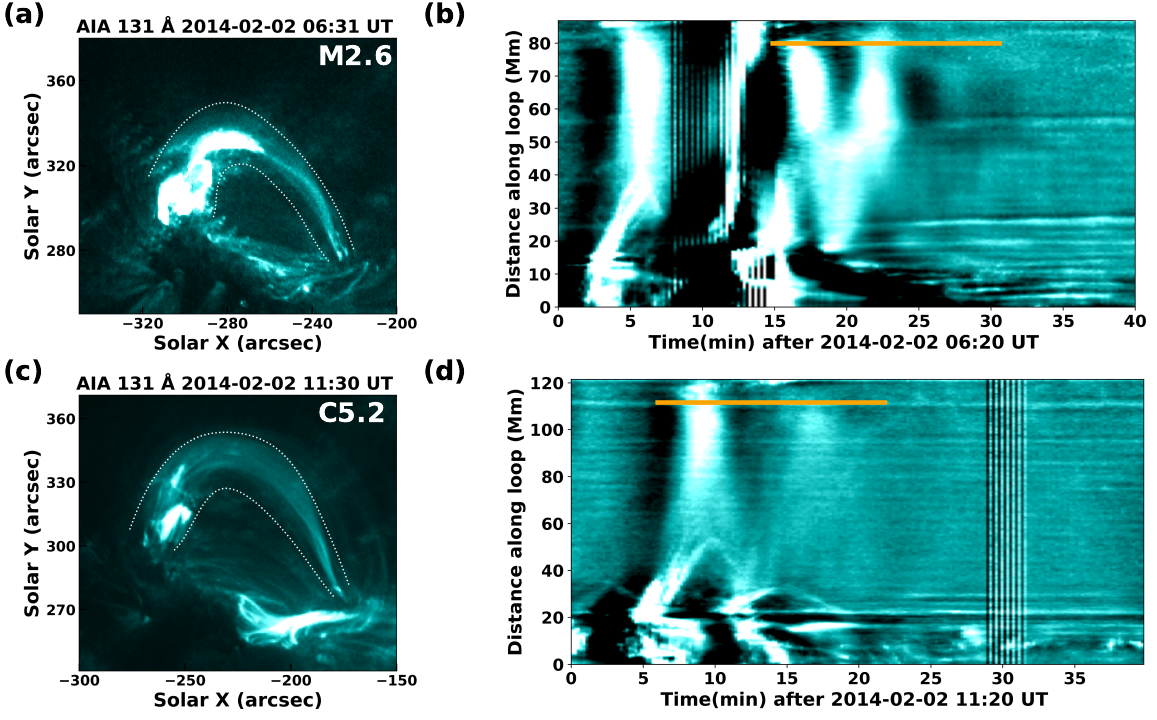}
   \caption{Similar to figure~\ref{xt} but for Loop~4 in table~\ref{table}. Note that in panel b), the brightening near 06:25 UT is associated with a small flare that occurred prior to the M-class flare considered here. The vertical stripes visible in both time--distance maps originate from the alternating low-exposure images automatically acquired by AIA in specific channels during flares. Despite doing exposure time normalisation, the intensities in these frames did not return to the same level resulting in these alternate stripes. The stripes in panel d) appear to begin long after the peak of the flare, however, the alternate low-exposure frames in this case are triggered by another flare from a different active region.}
\end{figure}

\begin{figure}[!h]
    \centering
    \includegraphics[width=.7\textwidth]{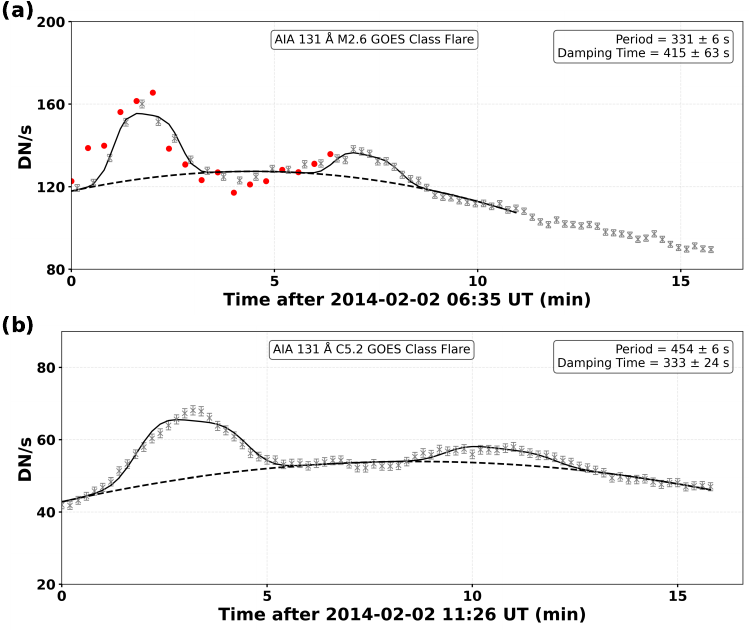}
   \caption{Similar to figure~\ref{fitting} but for Loop~4 in table~\ref{table}.}
\end{figure}

\begin{figure}[!h]
    \centering
    \includegraphics[width=1\textwidth]{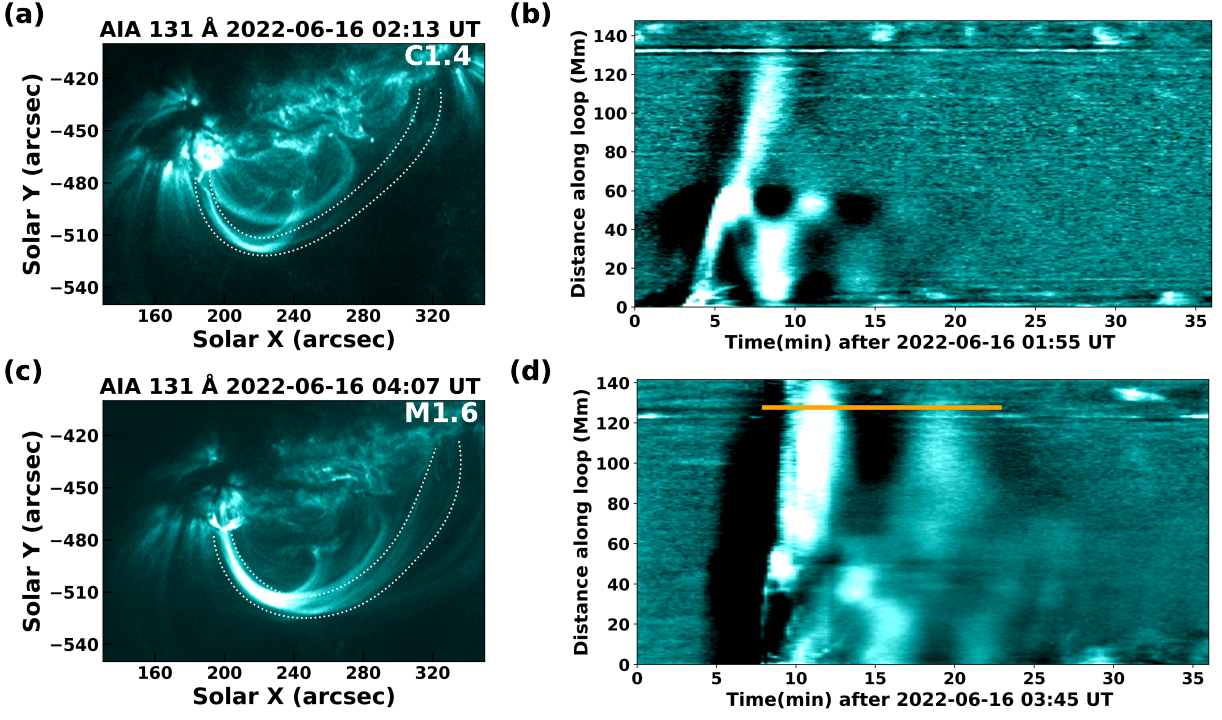}
   \caption{Similar to figure~\ref{xt} but for Loop~5 in table~\ref{table}. In the top-right panel, only a single bright ridge is visible in the time–distance map, preventing a reliable estimation of the oscillation properties. Consequently, no oscillation parameters were derived for that event. Note that the oscillatory pattern visible in the lower part of the time-distance map correspond to a different, smaller coronal loop along the line of sight visible in panel a) and analysed separately as Loop~3 (Table~\ref{table}).}
   \label{fig:fig12}
\end{figure}

\begin{figure}[!h]
    \centering
    \includegraphics[width=1\textwidth]{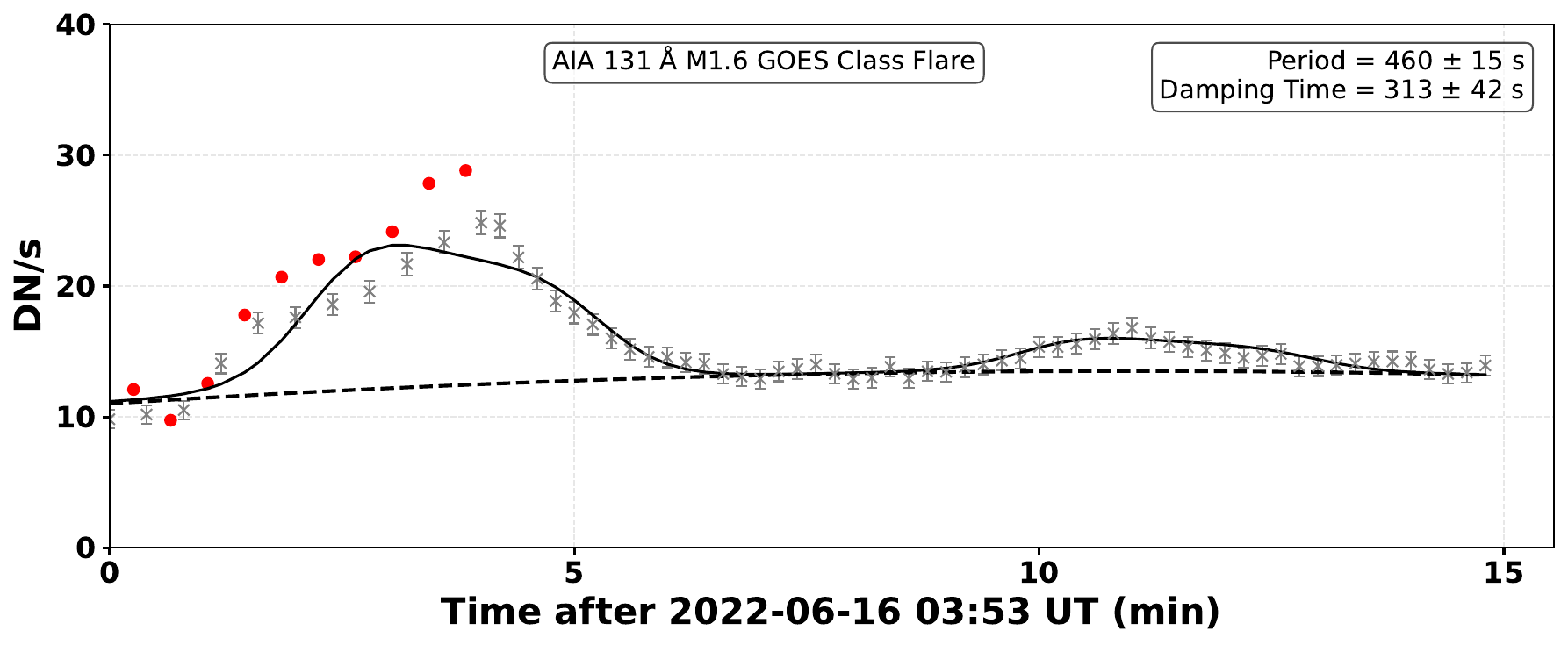}
   \caption{Same as Figure~\ref{fitting}, but for Event~b in Loop~5 listed in Table~\ref{table}.}
\end{figure}

\begin{figure}[!h]
    \centering
    \includegraphics[width=.8\textwidth]{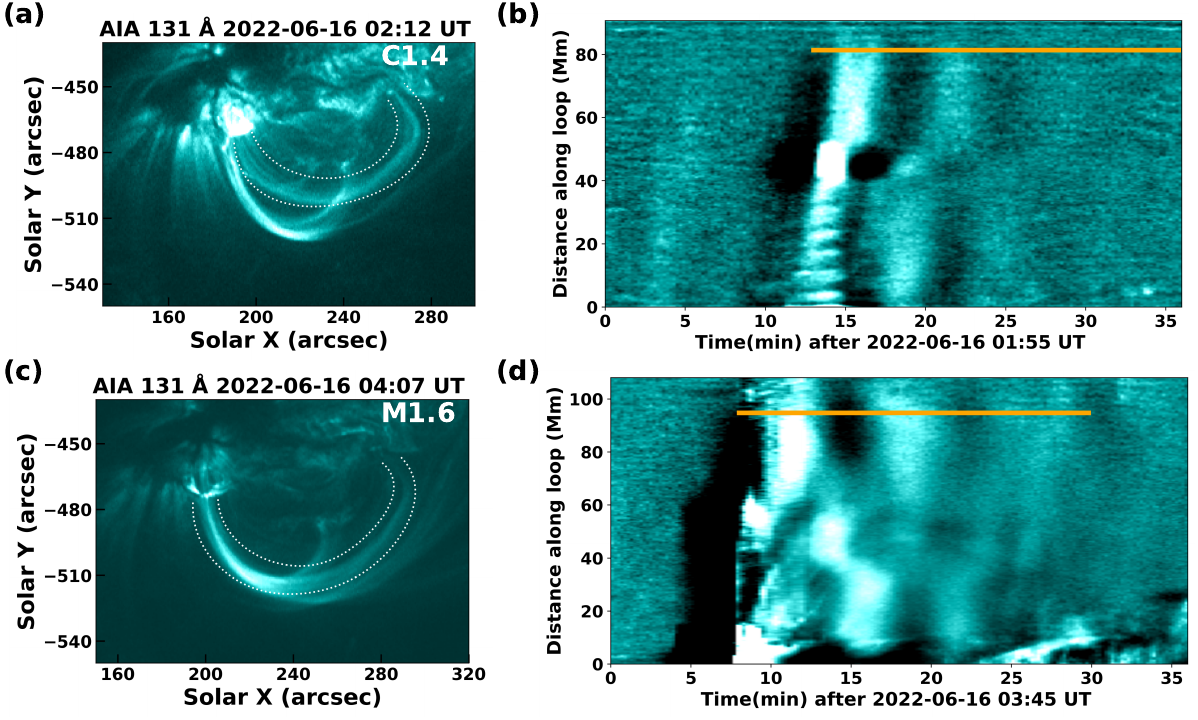}
   \caption{Similar to figure~\ref{xt} but for Loop~6 in table~\ref{table}.}
\end{figure}

\begin{figure}[!h]
    \centering
    \includegraphics[width=.7\textwidth]{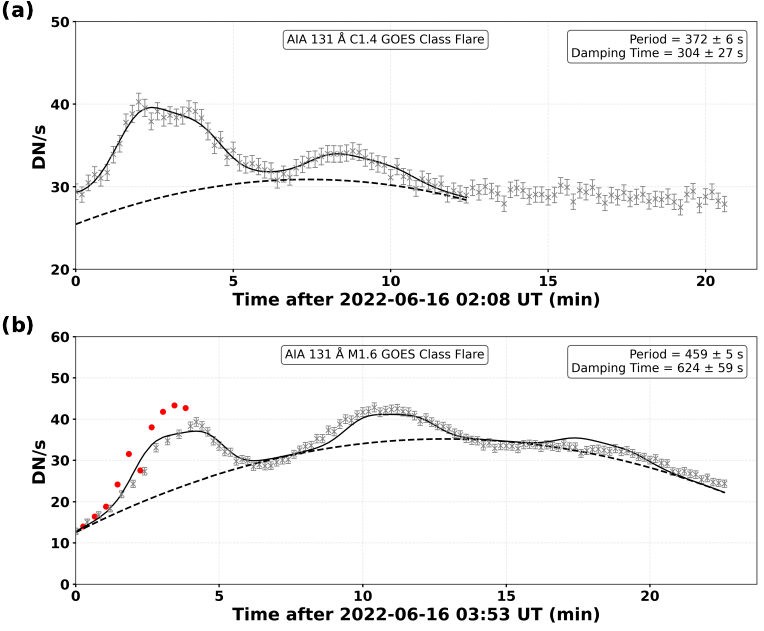}
   \caption{Similar to figure~\ref{fitting} but for Loop~6 in table~\ref{table}.}

\end{figure}

\begin{figure}[!h]
    \centering
    \includegraphics[width=.8\textwidth]{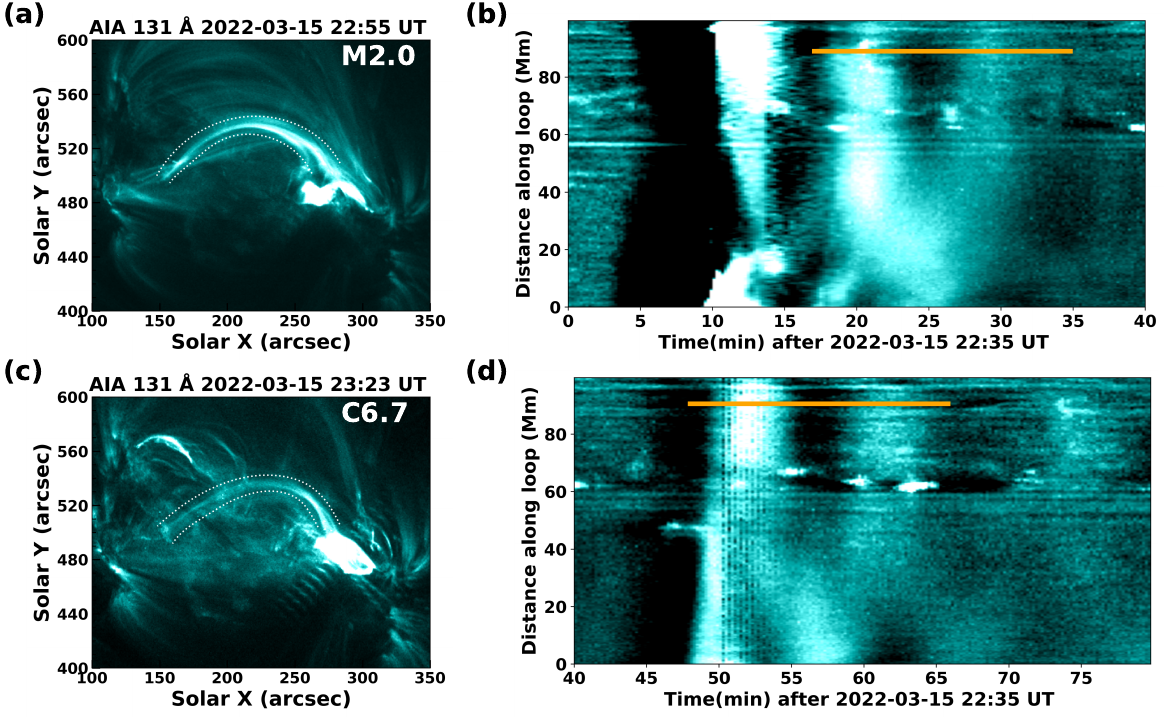}
   \caption{Similar to figure~\ref{xt} but for Loop~7 in table~\ref{table}.}
\end{figure}

\begin{figure}
    \centering
    \includegraphics[width=.7\textwidth]{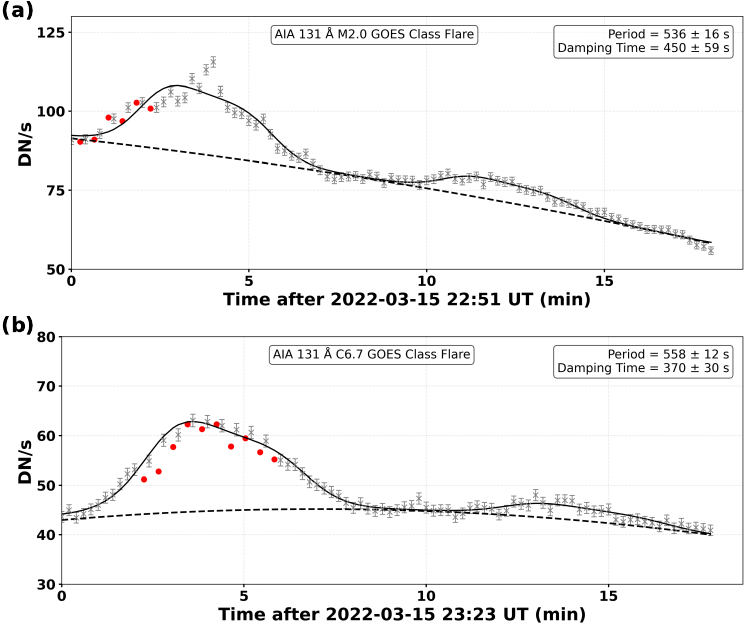}
   \caption{Similar to figure~\ref{fitting} but for Loop~7 in table~\ref{table}.}
\end{figure}

\bibliographystyle{aasjournalv7}



\end{document}